%% file: EsymAstro_LongMS.tex
\documentclass[aps,prc,superscriptaddress,twocolumn,showpacs,nofootinbib,preprintnumbers]{revtex4-1}

\usepackage[english]{babel}
\usepackage[T1]{fontenc}
\usepackage[colorlinks=true,linkcolor=blue,citecolor=blue,urlcolor=blue]{hyperref}
\usepackage{bm,bbm,amssymb,amsmath}
\usepackage{graphicx}
\usepackage{multirow,dcolumn,booktabs,enumerate}
\usepackage{isotope}
\usepackage{physics}
\usepackage{slashed}
\usepackage{nicefrac}
\usepackage[capitalize]{cleveref}
\usepackage[normalem]{ulem}

\usepackage{comment}
\usepackage{tikz}

\allowdisplaybreaks[1]

\newcommand{\fmiq}{\, \text{fm}^{-3}}

\newcommand{\mev}{\, \text{MeV}}

\include{EsymAstro_macros-final}

\begin{document}
\preprint{LA-UR-21-21363}

\title{
A Detailed Examination of Astrophysical Constraints on the Symmetry Energy and the Neutron Skin of $^{208}$Pb with Minimal Modeling Assumptions
}

\author{Reed Essick}
\email[E-mail:~]{reed.essick@gmail.com}
\affiliation{Perimeter Institute for Theoretical Physics, 31 Caroline Street North, Waterloo, Ontario, Canada, N2L 2Y5}
\affiliation{Kavli Institute for Cosmological Physics, The University of Chicago, Chicago, IL 60637, USA}
 
\author{Philippe Landry}
\email[E-mail:~]{plandry@fullerton.edu}
\affiliation{Nicholas \& Lee Begovich Center for Gravitational-Wave Physics \& Astronomy, California State University, Fullerton, 800 N State College Blvd, Fullerton, CA 92831}

\author{Achim Schwenk}
\email[E-mail:~]{schwenk@physik.tu-darmstadt.de}
\affiliation{Technische Universit\"at Darmstadt, Department of Physics, 64289 Darmstadt, Germany}
\affiliation{ExtreMe Matter Institute EMMI, GSI Helmholtzzentrum f\"ur Schwerionenforschung GmbH, 64291 Darmstadt, Germany}
\affiliation{Max-Planck-Institut f\"ur Kernphysik, Saupfercheckweg 1, 69117 Heidelberg, Germany}

\author{Ingo Tews}
\email[E-mail:~]{itews@lanl.gov}
\affiliation{Theoretical Division, Los Alamos National Laboratory, Los Alamos, NM 87545, USA}

\begin{abstract}
    The symmetry energy and its density dependence are pivotal for many nuclear physics and astrophysics applications, as they determine properties ranging from the neutron-skin thickness of nuclei to the crust thickness and the radius of neutron stars.
    Recently, PREX-II reported a value of \prexRskin~fm for the neutron-skin thickness of $^{208}$Pb, $R_{\rm skin}^{^{208}\text{Pb}}$, implying a symmetry-energy slope parameter $L$ of \prexL~MeV, larger than most ranges obtained from microscopic calculations and other nuclear experiments.
    We use a nonparametric equation of state representation based on Gaussian processes to constrain the symmetry energy $S_0$, $L$, and $R_{\rm skin}^{^{208}\text{Pb}}$ directly from observations of neutron stars with minimal modeling assumptions. 
    The resulting astrophysical constraints from heavy pulsar masses, LIGO/Virgo, and NICER favor smaller values of the neutron skin and $L$, as well as negative symmetry incompressibilities.
    Combining astrophysical data with chiral effective field theory (\EFT) and PREX-II constraints yields $S_0 = \mrgeftAstroPREXPostS$~MeV, $L=\mrgeftAstroPREXPostL$~MeV, and $R_{\rm skin}^{^{208}\text{Pb}} = \mrgeftAstroPREXPostRskin$~fm.
    We also examine the consistency of several individual \EFT~calculations with astrophysical observations and terrestrial experiments.
    We find that there is only mild tension between \EFT, astrophysical data, and PREX-II's $R_\mathrm{skin}^{^{208}\mathrm{Pb}}$ measurement ($p$-value $= \mrgeftAstroPostPREXpvalue$) and that there is excellent agreement between \EFT, astrophysical data, and other nuclear experiments.
\end{abstract}

\maketitle

\section{Introduction}
\label{sec:intro}

Knowledge of the nuclear symmetry energy is vital for describing systems with neutron-proton asymmetry, ranging from atomic nuclei to neutron stars~\cite{Lattimer:2012xj, Tsang:2012se, Huth:2020ozf}. 
The symmetry energy is defined as the difference between the nuclear energy per particle in pure neutron matter (PNM) and symmetric nuclear matter (SNM),
\begin{equation}
    S(n) = \frac{E_{\rm PNM}}{A}(n)-\frac{E_{\rm SNM}}{A}(n)\,.
    \label{eq:S0}
\end{equation}
Pure neutron matter consists only of neutrons and resembles neutron-star matter closely, while SNM consists of equal parts of protons and neutrons and can be probed through the bulk energy of atomic nuclei. 
The value of $S_0 = S(n_0)$, typically defined at nuclear saturation density $n_0\approx 0.16\, \mathrm{fm}^{-3}$, and the density dependence of $S(n)$, described by its slope parameter $L$ and curvature $K_{\rm sym}$, 
\begin{gather}
    L=3 n \left. \frac{\partial S(n)}{\partial n}\right|_{n_0} \,, \\
    K_{\rm sym}(n)=9 n^2 \left. \frac{\partial^2 S(n)}{\partial n^2}\right|_{n_0} \,,
     \label{eq:L-Ksym}
\end{gather}
can be correlated to several observables in nuclear physics and astrophysics, e.g., to the neutron-skin thickness of nuclei ($R_\mathrm{skin}$~\cite{Typel:2001, Vinas:2013hua, Reinhard:2016mdi, Mondal2016}), their electric dipole polarizability ($\alpha_D$~\cite{Tamii:2011pv, Piekarewicz:2012pp, RocaMaza:2013, RocaMaza:2015}), the radius ($R$) of neutron stars (NSs)~\cite{Lattimer:2000nx, Steiner:2012}, and properties of the NS crust~\cite{Neill:2020szr}.
This is because $L$ is related to the pressure of PNM at $n_0$, where $d (E_\mathrm{SNM}/A)/dn = 0$. 
Typical values for $S_0$ and $L$ from nuclear experiments~\cite{Lattimer:2012xj, Tsang:2012se, Tamii:2011pv, RocaMaza:2015, Russotto:2016} and theory~\cite{Huth:2020ozf, Tews:2012fj, Hebeler:2013nza, Drischler:2015eba, Lonardoni:2019ypg, Drischler:2020hwi} are $30$--$35$~MeV and $30$--$70$~MeV, respectively. 

In particular, the neutron-skin thickness of $^{208}$Pb, $R_\mathrm{skin}^{^{208}\mathrm{Pb}}$, is strongly correlated with $L$~\cite{Typel:2001, Vinas:2013hua, Reinhard:2016mdi, Mondal2016}.
Recently, the PREX collaboration determined $R_\mathrm{skin}^{^{208}\mathrm{Pb}}$ by measuring the parity-violating asymmetry ($\rm A_{\rm PV}$) in the elastic scattering of polarized electrons off $^{208}\mathrm{Pb}$. 
Using data from two experimental runs, PREX-I and PREX-II, the PREX collaboration reported $R_\mathrm{skin}^{^{208}\mathrm{Pb}} = \prexRskin\,\mathrm{fm}$ (mean $\pm$ standard deviation)~\cite{PREXII}.
Using a correlation between $R_\mathrm{skin}^{^{208}\mathrm{Pb}}$ and $L$, Ref.~\cite{Reed:2021nqk} inferred $L =\prexL\,\mathrm{MeV}$ from this measurement.
Note that Ref.~\cite{Reinhard:2021utv} has found lower values of $R_\mathrm{skin}^{^{208}\mathrm{Pb}}$ and $L$ when folding in information from other nuclear observables.

In recent work~\cite{Essick:2021kjb}, we examined astrophysical constraints on the symmetry energy, its density dependence, and $R_\mathrm{skin}^{^{208}\mathrm{Pb}}$ using a nonparametric inference framework for the equation of state (EOS)~\cite{LandryEssick2019, EssickLandryHolz2020}.
This framework is based on Gaussian Processes (GPs) that simultaneously represent the uncertainty in the (infinitely many) functional degrees of freedom of the sound speed in $\beta$-equilibrium as a function of pressure.
This approach avoids the modeling assumptions implicit in parametrized EOS representations---e.g., speed-of-sound~\cite{Alford:2013aca, Tews:2018chv, Greif:2018njt}, polytropic~\cite{Hebeler:2013nza, RaithelOzel2016}, or spectral~\cite{Lindblom:2010bb, Lindblom:2012zi} extension schemes---which attempt to capture the variability in the EOS in terms of a number of parameters.
Hence, our extraction of the symmetry energy and the neutron-skin thickness allows for increased model freedom relative to astrophysical inferences using explicit parameterizations of the EOS (e.g., Refs.~\cite{AlamAgrawal2016, CarsonSteiner2019, BiswasChar2020, Yue:2021yfx}).
Indeed, our approach reduces systematic uncertainties from \textit{a priori} modeling assumptions, which can otherwise be difficult to quantify, and provides constraints obtained directly from the astrophysical data.

In this paper, we provide a more detailed description of our method and present additional new results for symmetry-energy parameters, the neutron-skin thickness and NS properties.
In Ref.~\cite{Essick:2021kjb}, we marginalized over four nuclear-theory calculations of the EOS from chiral effective field theory ($\chi$EFT). 
Here, we examine the results for the individual calculations and discuss what we can learn about nuclear interactions from comparisons with astrophysical data.
In general, we find no significant tension between the PREX-II data and astrophysical observations, primarily because $L$ is less strongly correlated with NS observables than has typically been claimed~\cite{Lattimer:2000nx,Lattimer:2012xj}.
Given current measurement uncertainties, there is only mild tension between PREX-II and the \EFT~predictions, while the latter agree very well with measurements of the dipole polarizability of $^{208}$Pb ($\alpha_D^{^{208}\mathrm{Pb}}$)~\cite{Tamii:2011pv, RocaMaza:2013, RocaMaza:2015}.
Finally, we show that allowing for a nonparametric high-density extension of the EOS leads to a significantly weaker correlation of the $L$ parameter with NS radii, which must be taken into account when discussing the impact of a  precise $R_\mathrm{skin}^{^{208}\mathrm{Pb}}$ measurement on NS radii.

This paper is structured as follows. 
In Sec.~\ref{sec:nonparametric inference}, we introduce the nonparametric EOS inference scheme.
In Sec.~\ref{sec:esym}, we explain how we extract the nuclear parameters from the nonparametric EOS realizations.
We then present the results of the inference of microscopic and macroscopic dense-matter properties in Sec.~\ref{sec:results}. 
In particular, we address the consistency of various $\chi$EFT predictions with astrophysical observations and experimental $R_\mathrm{skin}^{^{208}\mathrm{Pb}}$ and $\alpha_D^{^{208}\mathrm{Pb}}$ measurements.
In Sec.~\ref{sec:discussion}, we discuss possible future areas of improvement and their expected impact before concluding in Sec.~\ref{sec:summary}.

\section{Methodology}
\label{sec:nonparametric inference}

We briefly review our GP-based nonparametric EOS inference scheme in Sec.~\ref{sec:nonparametric gaussian processes} before summarizing the astrophysical data used in our inference in Sec.~\ref{sec:astrophysical data}.
Section~\ref{sec:chiral eft calculations} describes the \EFT~calculations employed in this work, against which we contrast the constraints obtained without nuclear-theory input at low densities.

\subsection{Nonparametric EOS Inference}
\label{sec:nonparametric gaussian processes}

To extract dense matter information from astrophysical observations of NSs, we need a model for the NS EOS, i.e., the relation between energy density and pressure in the stellar interior. 
In this work, we use the nonparametric representation of the EOS introduced in Refs.~\cite{LandryEssick2019, EssickLandryHolz2020} based on GPs that model the uncertainty in the correlations between the sound speed in $\beta$-equilibrium at different pressures.
By construction, the GPs generate EOS realizations that are causal, thermodynamically stable, and matched to a NS crust model (BPS~\cite{BaymPethick1971}) at very low densities, $n< 0.3 n_0$.
Although GPs can be constructed to closely emulate the behavior of specific theoretical models, we instead construct GPs that explore as much functional behavior as possible (see the discussion of \emph{model-informed} vs. \emph{model-agnostic} priors in Refs.~\cite{LandryEssick2019, EssickLandryHolz2020}).
That is, our GPs are not strongly informed by a specific description of the microphysics; they are designed to be theory-agnostic.

Our GPs are conditioned on a training set of tabulated EOSs from the literature.
In particular, we follow Ref.~\cite{EssickLandryHolz2020} and construct priors from mixture models of GPs separately conditioned on hadronic, hyperonic and quark EOSs.
We condition 50 GPs with agnostic hyperparameters for each composition, and then marginalize over the compositions to obtain our final prior; see Ref.~\cite{EssickLandryHolz2020} for more details.
In this way, our prior emulates the functional behavior of established EOSs on average.
However, each process's uncertainties are very large, so that the EOS realizations we generate span a much wider range of behavior than the training set.
This includes EOSs that are much stiffer or much softer than EOSs from the literature, as well as many that exhibit sharp features reminiscent of strong phase transitions that can give rise to multiple stable branches in the mass-radius relation. 
By sampling many EOS realizations from the GPs, one obtains a discrete prior process over the EOS.
We typically draw $10^4$--$10^6$ EOS realizations for each prior we consider.

Given this large set of EOS realizations, our analysis proceeds through a Monte-Carlo implementation of a hierarchical Bayesian inference. 
Every EOS from the prior is assigned a marginal likelihood from each astrophysical observation.
In turn, the likelihood for each observation is modeled as an optimized kernel density estimate (KDE), and we directly marginalize over nuisance parameters (e.g., the masses $M$) with respect to a fixed prior (see Ref.~\cite{LandryEssickChatziioannou2020} for more details). 
This results in a representation of the posterior EOS process as a set of discrete samples with weights equal to the product of the marginal likelihoods.
The posterior probability for an EOS realization $\varepsilon_\beta$ is then
\begin{equation} \label{bayes}
    P(\varepsilon_\beta | \{d\}) \propto P(\varepsilon_\beta) \prod_i P(d_i | \varepsilon_\beta)\,,
\end{equation}
where $\{d\} = \{d_1,d_2,\dots\}$ is the set of observations, $P(d_i | \varepsilon_\beta)$ are the corresponding marginal likelihoods, and $P(\varepsilon_\beta)$ is the EOS realization's prior probability.

\subsection{Astrophysical Data}
\label{sec:astrophysical data}

The nonparametric inference scheme can incorporate different types of astrophysical observations~\cite{LandryEssickChatziioannou2020}, including the existence of massive pulsars~\cite{AntoniadisFreire2013, CromartieFonseca2020}, simultaneous $M$-$\Lambda$ measurements from compact binary mergers with gravitational waves (GWs)~\cite{gw170817, gw170817_props} observed by the Advanced LIGO~\cite{TheLIGOScientific:2014jea} and Virgo~\cite{TheVirgo:2014hva} interferometers, and simultaneous $M$-$R$ measurements from X-ray pulse-profile modeling of Neutron Star Interior Composition Explorer (NICER)~\cite{MillerLamb2019, Riley:2019yda} observations.
We use these astrophysical observations to constrain the GPs described in the previous section.

For the masses of the two heaviest known NSs, measured via pulsar timing, we model the likelihoods $P(d|m)$ as Gaussian distributions.
For PSR J0740+6620~\cite{CromartieFonseca2020, Fonseca2021} (respectively, PSR J0348+0432~\cite{AntoniadisFreire2013}) the mean and standard deviation are $2.08 \pm 0.07\,M_{\odot}$ ($2.01 \pm 0.04\,M_{\odot}$).
The likelihood of an EOS realization $\varepsilon_\beta$, given this observation, is
\begin{equation}
    P(d|\varepsilon_\beta) \propto \int P(d|M) P(M|\varepsilon_\beta) dM \,.
\end{equation}
We take the mass prior $P(M|\varepsilon_\beta)$ to be flat up to the maximum mass supported by the EOS realization, and take care to include the proper normalization. 
This ensures that EOSs that predict a maximum mass far below the pulsar mass are assigned zero likelihood, while among EOSs that support greater masses the 
models that least overestimate the maximum mass relative to the observation are favored (see Appendix of~\cite{LandryEssickChatziioannou2020} and discussion in Ref.~\cite{Miller2019}). In practical terms, this is because the non-observation of pulsars with masses significantly above $2.1\,M_\odot$ is informative in itself. 

For $M$-$\Lambda$ measurements from GW170817~\cite{gw170817, gw170817_props}, we model the likelihood $P(d|M_1,M_2,\Lambda_1,\Lambda_2)$ with an optimized Gaussian KDE as explained in Ref.~\cite{EssickLandryHolz2020}.
The corresponding likelihood of an EOS realization $\varepsilon_\beta$ given this observation is
\begin{multline}
    P(d|\varepsilon_\beta) \propto \int \Big{[} P(d|M_1,M_2,\Lambda_1,\Lambda_2) P(M_1,M_2) \\ \times \delta(\Lambda_1 - \Lambda(M_1)) \delta(\Lambda_2 - \Lambda(M_2)) \Big{]} dM_1 dM_2 \,.
\end{multline}
The mass prior is taken to be uniform.
We do not truncate it at the maximum mass supported by the EOS because we do not exclude \textit{a priori} the possibility that one of the components of the binary was a BH. 
Our analysis does not incorporate the binary NS observation GW190425, as it was not loud enough to yield a measurable matter signature and hence inform inference of the EOS.
Furthermore, we do not include light-curve models of electromagnetic counterparts associated with GW events because of the systematic uncertainties involved in interpreting the kilonova physics and its connection to the EOS (see, e.g., discussions in Refs.~\cite{gw170817_kilonova, Coughlin:2018miv, Kasen:2019, Siegel:2019mlp, Metzger:2019zeh, Dietrich:2020efo, Stachie:2021, Raaijmakers:2021uju}).

Finally, we consider X-ray pulse-profile measurements of PSR J0030+0451's mass and radius assuming a three-hotspot configuration~\cite{MillerLamb2019} (see also Ref.~\cite{Riley:2019yda}, which yields comparable results~\cite{LandryEssickChatziioannou2020}).
The likelihood $P(d|M,R)$ for this observation is also modeled with an optimized Gaussian KDE~\cite{EssickLandryHolz2020}.
Weighing an EOS realization $\varepsilon_\beta$ by this likelihood, we obtain
\begin{equation}
    P(d|\varepsilon_\beta) \propto \int P(d|M,R) P(M|\varepsilon_\beta) dM \,.
\end{equation}
The mass prior should, in principle, extend only up to the maximum mass for a given EOS realization because, like for the pulsar mass measurements, we know that PSR J0030+0451 is a NS.
However, for convenience we instead assume a NS population model that truncates the mass prior for X-ray sources well below the maximum NS mass.
As discussed in Ref.~\cite{LandryEssickChatziioannou2020}, these two prescriptions are effectively equivalent in the case of PSR J0030+0451 because its mass is clearly smaller than the maximum mass of any viable EOS.

Nonetheless, we would need to truncate $P(M|\varepsilon_\beta)$ at $M_\mathrm{max}$ if we were to include the recent NICER+XMM Newton observations of J0740+6620~\cite{Miller:2021qha, Riley:2021pdl, Raaijmakers:2021uju}. 
We do not consider this measurement in the present work because the NICER results for J0740+6620 were published after Ref.~\cite{Essick:2021kjb} and the properties of this high-mass NS do not influence significantly the EOS inference at $n_0$ (see also Refs.~\cite{Pang:2021jta, Legred:2021hdx}), especially within our nonparametric framework (see, e.g., Fig.~\ref{fig:R(M=1.4), Lambda(M=1.4) and L}).
However, the updated mass measurement for J0740+6620 reported in Ref.~\cite{Fonseca2021} is incorporated as one of the two pulsar mass observations described above. 

\subsection{Chiral EFT Calculations}
\label{sec:chiral eft calculations}

The nonparametric EOS prior based on a crust EOS with GP extensions to higher densities can also be conditioned on theoretical calculations of the EOS for densities above the crust and up to around $1-2 n_0$, where nuclear theory calculations are well controlled.
At higher densities, our EOS framework still uses the model independence of the GP construction~\cite{EssickTewsLandryReddyHolz2020}.
Following our previous work~\cite{Essick:2021kjb}, we separately condition the EOS on the uncertainty band obtained from four different calculations based on $\chi$EFT interactions and marginalize over all four bands.

First, we consider quantum Monte Carlo calculations (QMC) using local $\chi$EFT interactions up to next-to-next-to-leading order (N$^2$LO)~\cite{Lynn:2016}.
These results, labeled \QMC, are based on a nonperturbative many-body method that is proven to be accurate for strongly correlated systems, but are presently limited to N$^2$LO due to nonlocalities entering at higher order in $\chi$EFT.
As a result, the \QMC~band has somewhat larger uncertainties.
In addition, we consider two calculations based on many-body perturbation theory (MBPT)~\cite{Tews:2012fj,Drischler:2017wtt} using nonlocal $\chi$EFT interactions up to next-to-next-to-next-to-leading order (N$^3$LO).
Both calculations include all two-, three-, and four-neutron interactions up to this order.
The results from Ref.~\cite{Drischler:2017wtt}, which we label MBPT$^\text{(2019)}_\text{N$^3$LO}$, include contributions up to higher order in MBPT as well as EFT truncation uncertainties (for two cutoffs: 450 and 500\,MeV), while the results from Ref.~\cite{Tews:2012fj}, labeled MBPT$^\text{(2013)}_\text{N$^3$LO}$, are lower order in MBPT but include other uncertainties in two- and three-nucleon interactions as well.
Therefore, we find it useful to explore both EOS bands here.
We note that the combined 450 and 500\,MeV N$^3$LO bands from Ref.~\cite{Drischler:2017wtt} overlap very closely with the recent GP uncertainty bands (GP-B) from Ref.~\cite{Drischler:2020hwi}, labeled \DrischlerGP in the following (see also Ref.~\cite{Huth:2020ozf}).
Finally, we also consider the MBPT calculations with two-nucleon interactions at N$^3$LO and three-nucleon interactions at N$^2$LO, labeled MBPT$^\text{(2010)}_\text{mixed}$, based on a broader range of three-nucleon couplings~\cite{Hebeler:2009iv,Hebeler:2013nza}.
Exploring these four bands allows us to account for different nuclear interactions and many-body approaches, increasing the robustness of our results.

\section{Extraction of Nuclear Parameters from Nonparametric EOS Realizations}
\label{sec:esym}

The nuclear EOS can be described by the nucleonic energy per particle, $E_{\rm nuc}/A(n, x)$, which depends on the density $n$ and the proton fraction $x=n_p/n$ with $n_p$ being the proton density. 
The symmetry energy $S(n)$ is encoded in the $x$ dependence of $E_{\rm nuc}/A(n, x)$.
In our approach, we approximate the $x$ dependence of the nucleonic energy per particle with the standard quadratic expansion, 
\begin{equation}
    \frac{E_{\rm nuc}}{A}(n,x) = \frac{E_{\rm SNM}}{A} + S(n) (1-2x)^2  \label{eq:EOS}\,,
\end{equation}
where higher-order terms beyond $\mathcal{O}(x^2)$ are expected to be small around $n_0$, and can be safely neglected given current EOS uncertainties~\cite{Drischler:2013iza, Somasundaram:2020chb}.
For example, Ref.~\cite{Goriely2013} suggested systematic shifts of $\mathcal{O}(3\,\mathrm{MeV})$ in $L$ when higher-order terms are included in Eq.~\eqref{eq:EOS} (compare $L$ and $\tilde L$ in Table V), but these are much smaller than the statistical uncertainty in all our priors (Table~\ref{tab:credible regions}).
$S(n)$ can be computed as
\begin{align}
    S(n)&=\frac{E_{\rm nuc}}{A}\left(n, 0\right)-\frac{E_{\rm nuc}}{A}\left(n, 1/2 \right) \nonumber \\
    &= \frac{E_{\rm PNM}}{A}-\frac{E_{\rm SNM}}{A}\,. \label{eq:Sn}
\end{align}

In our nonparametric EOS inference, each EOS realization is represented in terms of the baryon density $n$, the energy density $\varepsilon_{\beta}$, and the pressure $p_{\beta}$ in $\beta$-equilibrium. 
These quantities are related to the energy per particle $E/A$ through
\begin{align}
    \varepsilon &= n\cdot \left(\frac{E}{A} +m_{\rm N}  \right) \,, \\ 
    p &= n^2 \frac{\partial E/A}{\partial n} \,,
\end{align}
where $m_N$ is the average nucleon mass and we use units with $\hbar = c = 1$.
We need to correct the total energy density in $\beta$-equilibrium for the contribution of electrons:
\begin{equation}
    \frac{E_{\rm nuc}}{A}(n,x) = \frac{\varepsilon_\beta(n)-\varepsilon_{\rm e}(n,x)}{n} - m_{\rm N}\,. \label{eq:Enuc}
\end{equation}
In this work, we describe the electron contribution using the relations for a relativistic Fermi gas~\cite{Chamel:2008ca}: 
\begin{multline}
    \varepsilon_{e}(n_{\rm e}) = \frac{m_{\rm e}^4}{8 \pi^2}
    \left(x_r(2 x_r^2+1)\sqrt{x_r^2+1}\right. \\ 
    \left.-\ln(x_r+\sqrt{x_r^2+1})\right)\,.
    \label{eq:el_eps}
\end{multline}
where $n_e$ is the electron density, and $x_r= k_F / m_{\rm e} =(3 \pi^2 n_{\rm e})^{1/3}/m_{e}$ with the electron mass $m_{\rm e}=0.511$\,MeV.
We neglect the contribution from muons because their effect on the EOS around nuclear saturation density is small.
Then, due to charge neutrality, the electron density in $\beta$-equilibrium equals the proton density, $n_{\rm e} = x(n) \cdot n$.

The proton fraction $x(n)$ is unknown for each EOS draw but it can be constrained by enforcing the $\beta$-equilibrium condition,
\begin{equation}
    \mu_{\rm n}(n,x)=\mu_{\rm p}(n,x)+\mu_{\rm e}(n,x)\,,
    \label{eq:beta_equi}
\end{equation}
where $\mu_{\rm i}(n,x)$ is the chemical potential for particle species $i$.
The electron chemical potential is given by 
\begin{equation}
    \mu_{\rm e}(n_{\rm e}) = \sqrt{ (3 \pi^2 n_{\rm e})^{2/3} + m_{\rm e}^2 }\,,
    \label{eq:el_mu}
\end{equation}
and the neutron and proton chemical potentials $\mu_{\rm n}$ and $\mu_{\rm p}$ in asymmetric nuclear matter are given by
\begin{widetext}
    \begin{align}
        \mu_{\rm p}(n,x) & = \frac{d\varepsilon_\mathrm{nuc}}{d n_p} = n \frac{\partial \left(E_{\rm nuc}/A\right)}{\partial n} + \frac{\partial \left(E_{\rm nuc}/A\right)}{\partial x}(1-x)+\frac{E_{\rm nuc}}{A}+m_{\rm p}\,, \\
        \mu_{\rm n}(n,x) & = \frac{d\varepsilon_\mathrm{nuc}}{d n_n} = n \frac{\partial \left(E_{\rm nuc}/A\right)}{\partial n} - \frac{\partial \left(E_{\rm nuc}/A\right)}{\partial x}x+\frac{E_{\rm nuc}}{A}+m_{\rm n}\,,\label{eq:nuc_mu}
    \end{align}
\end{widetext}
with the neutron and proton masses $m_{\rm n}$ and $m_{\rm p}$, respectively. 
Hence, the $\beta$-equilibrium condition is given by
\begin{equation}\label{eq:beta_sol}
    m_{\rm n}-m_{\rm p}-\frac{\partial \left(E_{\rm nuc} / A \right)}{\partial x}-\mu_{\rm e}(n,x)=0\,.
\end{equation}

From Eqs.~\eqref{eq:EOS} and \eqref{eq:Sn}, the derivative of the nucleonic energy per particle with respect to $x$ is given by
\begin{equation}
    \frac{\partial \left(E_{\rm nuc}/A\right)}{\partial x} =-4\left( \frac{E_{\rm PNM}}{A} - \frac{E_{\rm SNM}}{A}  \right) (1-2x)\,.
    \label{eq:deriv_x}
\end{equation}
For the energy per particle of SNM, we can employ the standard Taylor expansion about $n_0$, 
\begin{equation}
    \frac{E_{\rm SNM}}{A}(n)=E_0+\frac12 K_0 \left(\frac{n-n_0}{3n_0}\right)^2+\cdots\,,
     \label{eq:SNM}
\end{equation}
where $n_0$, the saturation energy $E_0$, and the incompressibility $K_0$ are constrained empirically.
Higher-order terms beyond $K_0$ can be neglected because we determine the symmetry energy only around $n_0$.
See the Supplemental Material in Ref.~\cite{Essick:2021kjb} for a quantification of the effect of higher-order terms in $n$ and the presence of muons near saturation density.
For the parameters $n_0$, $E_0$, and $K_0$, we use the ranges from Ref.~\cite{Huth:2020ozf} (means $\pm$ standard deviations of Gaussian distributions):
\begin{align}
    n_0 &= 0.164\pm 0.007 \fmiq\,, \nonumber\\
    E_0 &= -15.86\pm 0.57 \mev\,, \label{eq:SNMparameters}\\
    K_0 &= 215\pm 40\mev \,.\nonumber
\end{align}

Putting all of this together, $\beta$-equilibrium must satisfy
\begin{multline}
    \frac{1-2x_{\beta}}{4}\Bigl( m_p - m_n + \mu_\mathrm{e}(n,x_{\beta})\Bigr) = \\
    \left( \frac{\varepsilon_\beta - \varepsilon_\mathrm{e}(n,x_{\beta})}{n} - m_\mathrm{N} - \frac{E_{\rm SNM}}{A}(n) \right)\,.
    \label{eq:fbeta}
\end{multline}
We self-consistently reconstruct the proton fraction for each EOS realization by solving Eq.~\eqref{eq:fbeta} for $x_{\beta}$ as a function of $n$ around $n_0$.
For this, we draw the parameters $E_0$, $K_0$, and $n_0$ from their empirical distributions in Eq.~\eqref{eq:SNMparameters} separately for each EOS, thereby marginalizing over their uncertainty within our Monte-Carlo sums over EOS realizations. 
We then calculate the PNM energy per particle $E_{\rm PNM}/A(n)$, the symmetry energy $S_0$, its derivative $L$, and its curvature $K_{\rm sym}$ as a function of baryon density $n$ in the vicinity of $n_0$ and report their values at the reference density,  $n_0^{\mathrm{ref}}=0.16\,\mathrm{fm}^{-3}$.
In the following we use $n_0$ to denote this reference density, but note again that the uncertainty in the empirical saturation point, Eq.~(\ref{eq:SNMparameters}), is included when extracting $S_0$, $L$, and $K_\mathrm{sym}$ from EOS samples.

\begin{figure}
    \centering
    \includegraphics[width=\columnwidth]{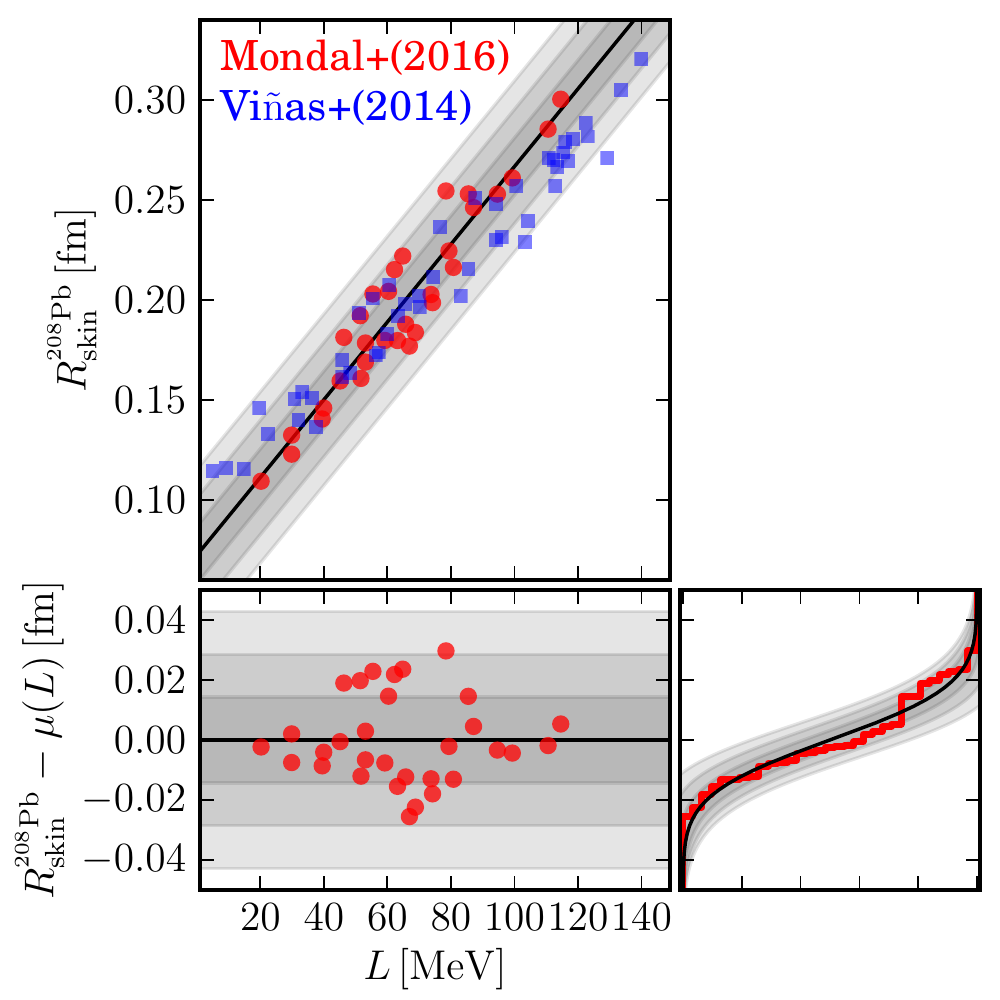}
    \caption{
        Uncertainty relation between $R_\mathrm{skin}^{^{208}\mathrm{Pb}}$ and $L$ modeled on the \externalresult{31} models from Ref.~\cite{Mondal2016} (\emph{red circles}) compared with  \externalresult{47} models from Ref.~\cite{Vinas:2013hua} (\emph{blue squares}).
        (\emph{left}) We model the theoretical uncertainty with a conditional probability $P(R_\mathrm{skin}^{^{208}\mathrm{Pb}}|L)$ using a normal distribution with mean given by Eq.~\eqref{eq:L2Rskin}.
        Shaded bands correspond to 1, 2, and 3-$\sigma$ uncertainties for $R_\mathrm{skin}^{^{208}\mathrm{Pb}}$ at each $L$.
        (\emph{bottom right}) Predicted cumulative distribution of residuals and empirical distribution based on the fit to Ref.~\cite{Mondal2016}, showing good quantitative agreement between our model and the scatter between the theoretical calculations.
        We note that the models from Ref.~\cite{Vinas:2013hua} are systematically shifted compared to Ref.~\cite{Mondal2016}, but they are well represented by our uncertainty model.
        }
    \label{fig:L2Rskin}
\end{figure}

\begin{figure}
    \centering
    \includegraphics[width=\columnwidth]{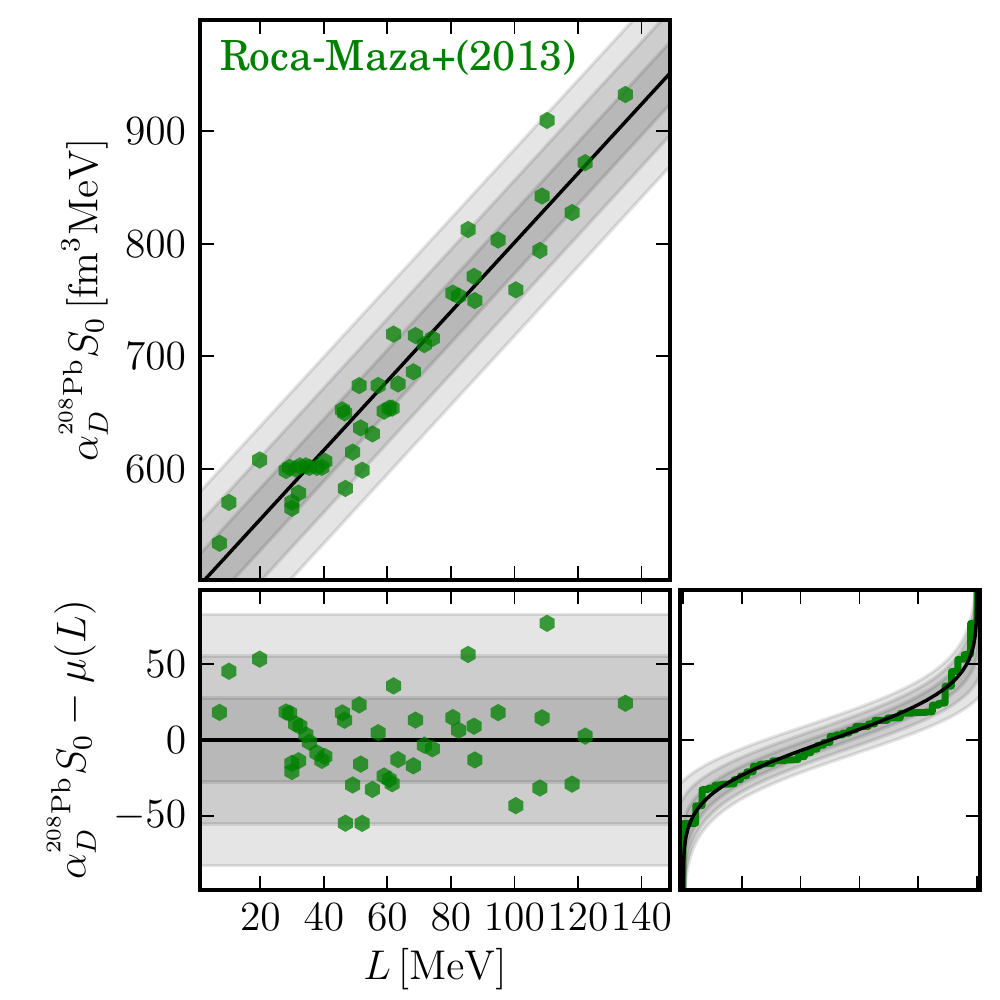}
    \caption{
        Analogous to Fig.~\ref{fig:L2Rskin}, but showing the conditional uncertainty for $P(\alpha_D^{^{208}\mathrm{Pb}} S_0|L)$, modeled as a Gaussian with mean given by Eq.~\eqref{eq:L2alphDS0}, based on Ref.~\cite{RocaMaza:2013}.
        Shaded bands represent 1, 2, and 3-$\sigma$ uncertainty within our model.
        We again obtain good quantitative agreement between our uncertainty model and the observed scatter of the theoretical models.
    }
    \label{fig:L2alphaD}
\end{figure}

With the mapping between the EOS and the parameters $E_\mathrm{PNM}/A$, $S_0$, $L$, and $K_\mathrm{sym}$ established, we calculate a posterior distribution
\begin{multline}
    P(E_\mathrm{PNM}/A, S_0, L, K_{\rm sym} | \{d\}) = \\
    \int \mathcal{D}\varepsilon_\beta\, P(\varepsilon_\beta | \{d\}) P(E_\mathrm{PNM}/A, S_0, L, K_{\rm sym} | \varepsilon_\beta)
\end{multline}
over the nuclear physics properties by conditioning on the astrophysical observations and marginalizing over many EOS realizations.

To extract the neutron-skin thickness of $^{208}$Pb, we use an empirical fit between $R_\mathrm{skin}^{^{208}\mathrm{Pb}}$ and $L$ based on the data in Ref.~\cite{Mondal2016}:
\begin{equation}\label{eq:L2Rskin}
    R_\mathrm{skin}^{^{208}\mathrm{Pb}}\,[\mathrm{fm}] = \result{0.0724 + 0.0019 \times (L\,[\mathrm{MeV}])} .
\end{equation}
This fit is calculated from a range of nonrelativistic Skyrme and relativistic energy-density functionals.
To model the uncertainty in this empirical relation, we fit the distribution of ($R_\mathrm{skin}^{^{208}\mathrm{Pb}}$, $L$) from Ref.~\cite{Mondal2016} to a Gaussian with a mean given by Eq.~\eqref{eq:L2Rskin}, obtaining a standard deviation of \result{$0.0143\,\mathrm{fm}$}.
This uncertainty model and the residuals of the fit are shown in Fig.~\ref{fig:L2Rskin}.
We also compare this fit with the density functionals used in Ref.~\cite{Vinas:2013hua}. Our fit provides a good representation of the spread between all these models.

Similarly, to connect our results to the electric dipole polarizability of $^{208}$Pb,  $\alpha_D^{^{208}\mathrm{Pb}}$, we use an empirical fit between $\alpha_D^{^{208}\mathrm{Pb}} \cdot S_0$ and $L$ based on Ref.~\cite{RocaMaza:2013}, finding
\begin{equation}\label{eq:L2alphDS0}
    \alpha_D^{^{208}\mathrm{Pb}} \cdot S_0\,[\mathrm{fm}^3\mathrm{MeV}]= \result{493.5 + 3.08 \times (L\,[\mathrm{MeV}])} \,.
\end{equation}
We again model the conditional distribution $P(\alpha_D^{^{208}\mathrm{Pb}} S_0|L)$ as a Gaussian with mean given by Eq.~\eqref{eq:L2alphDS0} and a standard deviation of \result{$27.6\,\mathrm{fm}^3\mathrm{MeV}$}.
This uncertainty model is shown in Fig.~\ref{fig:L2alphaD}.

\section{Results}
\label{sec:results}

\begin{figure*}[t]
    \begin{tikzpicture}
        \node (main) {\includegraphics[width=1.0\textwidth]{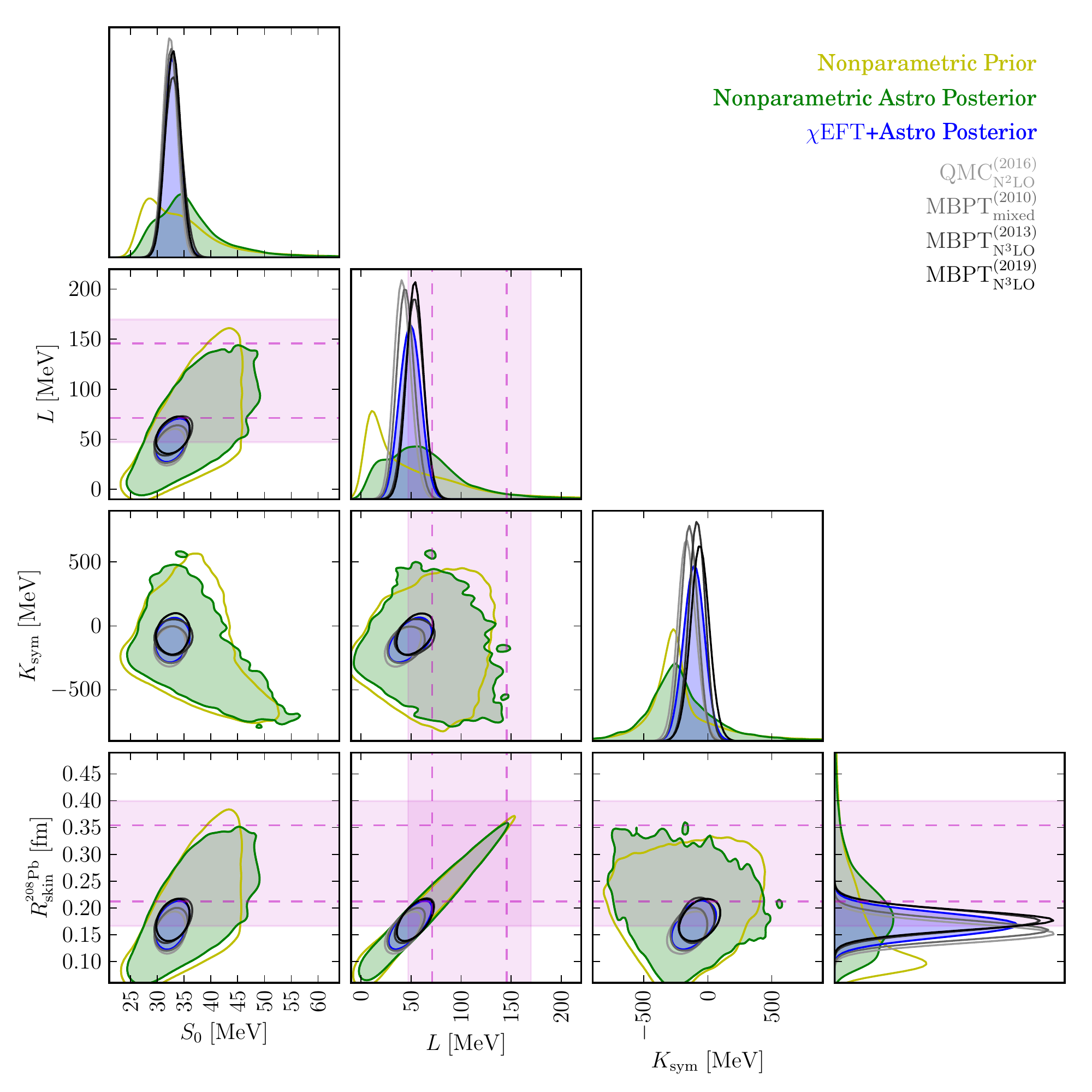}} ;
        \node (legend) [xshift=-0.50cm, yshift=+6.65cm] {\includegraphics[width=0.30\textwidth, clip=True, trim=13.0cm 14.75cm 0.75cm 0.75cm]{kde-corner-samples_decomposed-SLKRskin-Mondalfit.pdf}} ;
        \node (inset) [xshift=+5.70cm, yshift=+4.75cm] {\includegraphics[width=0.40\textwidth]{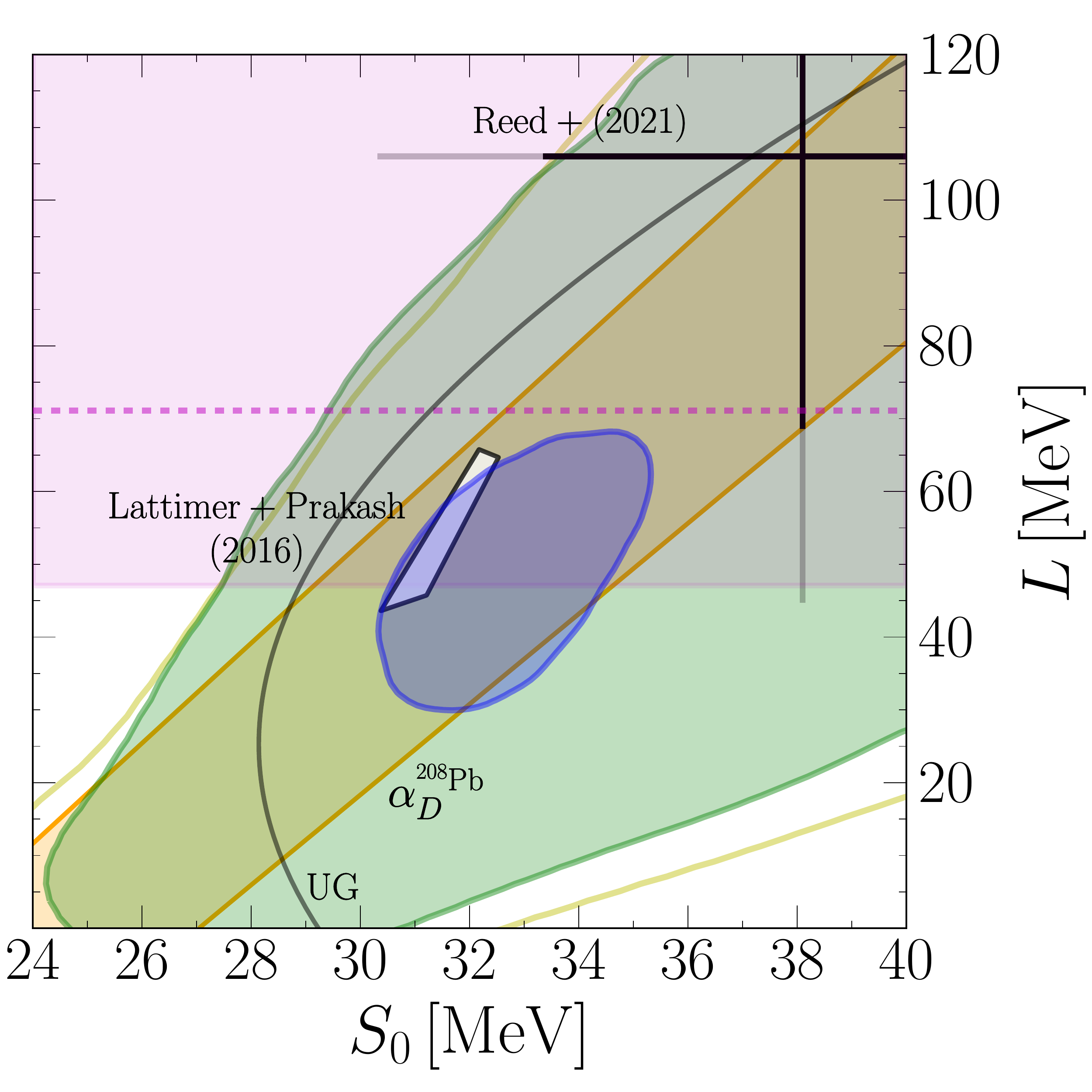}} ;
        \node (label) [right of=inset, xshift=-3.50cm, yshift=+2.50cm] {\textcolor{red}{PREX-II}} ;
    \end{tikzpicture}
    \vspace*{-10mm}
    \caption{
        Correlations between $S_0$, $L$, $K_{\rm sym}$, and $R_\mathrm{skin}^{^{208}{\rm Pb}}$ within our nonparametric prior (\emph{unshaded yellow}) and Astro-only posterior (\emph{shaded green}) as well as the \EFT-marginalized (\emph{shaded blue}), \QMC, \Hebeler, \MBPT, and \DrischlerPRL~Astro-only posteriors (\emph{unshaded greys}, ordered from lighter to darker with increasing $L$, see Table~\ref{tab:credible regions}).
        Joint distributions show 90\% credible regions, and the horizontal bands (\emph{pink}) represent PREX-II 90\% credible regions, with dashed lines the corresponding 68\% (1-$\sigma$) regions.
        The expanded ($S_0$, $L$) panel (\emph{upper right}) compares our nonparametric prior, Astro-only posterior, and \EFT+Astro posterior to other constraints: (\emph{white region}) Lattimer and Prakash~\cite{LATTIMER2016127} (overlap region of various nuclear experimental constraints), the unitary-gas (UG) bound from Ref.~\cite{Tews:2016jhi}, and the values reported by Reed \textit{et al.}~\cite{Reed:2021nqk} based on the PREX-II results.
        In addition, we show the correlation obtained from the experimental  $\alpha_D^{^{208}\mathrm{Pb}}$~\cite{Tamii:2011pv} using Eq.~\eqref{eq:L2alphDS0}.
    }
    \label{fig:decomposed esym corner}
\end{figure*}

\begin{table*}
    \centering
    {\renewcommand{\arraystretch}{1.4}
    \begin{tabular}{@{\extracolsep{0.08cm}} l l c c c c c c}
        \hline
          & &  $\frac{E_\mathrm{PNM}}{A}(n_0)\ [\mathrm{MeV}]$ &  $S_0\ [\mathrm{MeV}]$ & $L\ [\mathrm{MeV}]$ & $K_\mathrm{sym}\ [\mathrm{MeV}]$ & $R_\mathrm{skin}^{^{208}\mathrm{Pb}}\ [\mathrm{fm}]$ & $\alpha_D^{^{208}\mathrm{Pb}}\ [\mathrm{fm}^3]$ \\
        \hline
        \hline
        \multirow{4}{*}{Nonparametric}
         & Prior
            & \mrgagnPriorEpnm & \mrgagnPriorS & \mrgagnPriorL & \mrgagnPriorKsym & \mrgagnPriorRskin & \mrgagnPriorAlphaD \\
         & Astro Posterior
            & \mrgagnAstroPostEpnm & \mrgagnAstroPostS & \mrgagnAstroPostL & \mrgagnAstroPostKsym & \mrgagnAstroPostRskin & \mrgagnAstroPostAlphaD \\
         & Astro+PREX-II Post.
            & \mrgagnAstroPREXPostEpnm & \mrgagnAstroPREXPostS & \mrgagnAstroPREXPostL & \mrgagnAstroPREXPostKsym & \mrgagnAstroPREXPostRskin & \mrgagnAstroPREXPostAlphaD \\
         & Astro+$\alpha_D^{^{208}\mathrm{Pb}}$ Post.
            & \mrgagnAstroAlphaDPostEpnm & \mrgagnAstroAlphaDPostS & \mrgagnAstroAlphaDPostL & \mrgagnAstroAlphaDPostKsym & \mrgagnAstroAlphaDPostRskin & \mrgagnAstroAlphaDPostAlphaD \\
        \hline
        \hline
        \multirow{4}{*}{\EFT-marginalized}
         & Prior
            & \mrgeftPriorEpnm & \mrgeftPriorS & \mrgeftPriorL & \mrgeftPriorKsym & \mrgeftPriorRskin & \mrgeftPriorAlphaD \\
         & Astro Posterior
            & \mrgeftAstroPostEpnm & \mrgeftAstroPostS & \mrgeftAstroPostL & \mrgeftAstroPostKsym & \mrgeftAstroPostRskin & \mrgeftAstroPostAlphaD \\
         & Astro+PREX-II Post.
            & \mrgeftAstroPREXPostEpnm & \mrgeftAstroPREXPostS & \mrgeftAstroPREXPostL & \mrgeftAstroPREXPostKsym & \mrgeftAstroPREXPostRskin & \mrgeftAstroPREXPostAlphaD \\
         & Astro+$\alpha_D^{^{208}\mathrm{Pb}}$ Post.
            & \mrgeftAstroAlphaDPostEpnm & \mrgeftAstroAlphaDPostS & \mrgeftAstroAlphaDPostL & \mrgeftAstroAlphaDPostKsym & \mrgeftAstroAlphaDPostRskin & \mrgeftAstroAlphaDPostAlphaD \\
        \hline
        \multirow{5}{*}{\QMC~\cite{Lynn:2016}}
         & Original Work & \externalresult{[14.2, 18.8]} & \externalresult{[28.6, 36.2]} & \externalresult{[23.8, 58.2]} & - & - & -\\
         & Prior
            & \qmcPriorEpnm & \qmcPriorS & \qmcPriorL & \qmcPriorKsym & \qmcPriorRskin & \qmcPriorAlphaD \\
         & Astro Posterior
            & \qmcAstroPostEpnm & \qmcAstroPostS & \qmcAstroPostL & \qmcAstroPostKsym & \qmcAstroPostRskin & \qmcAstroPostAlphaD \\
         & Astro+PREX-II Post.
            & \qmcAstroPREXPostEpnm & \qmcAstroPREXPostS & \qmcAstroPREXPostL & \qmcAstroPREXPostKsym & \qmcAstroPREXPostRskin & \qmcAstroPREXPostAlphaD \\
         & Astro+$\alpha_D^{^{208}\mathrm{Pb}}$ Post.
            & \qmcAstroAlphaDPostEpnm & \qmcAstroAlphaDPostS & \qmcAstroAlphaDPostL & \qmcAstroAlphaDPostKsym & \qmcAstroAlphaDPostRskin & \qmcAstroAlphaDPostAlphaD \\
        \cline{2-8}
        \multirow{5}{*}{\Hebeler~\cite{Hebeler:2009iv,Hebeler:2013nza}}
        & Original Work & \externalresult{[14.3, 18.4]} & \externalresult{[29.7, 33.2]} & \externalresult{[32.5, 57.0]} & - & \externalresult{[0.14, 0.20]} & - \\
         & Prior
            & \HebelerPriorEpnm & \HebelerPriorS & \HebelerPriorL & \HebelerPriorKsym & \HebelerPriorRskin & \HebelerPriorAlphaD \\
         & Astro Posterior
            & \HebelerAstroPostEpnm & \HebelerAstroPostS & \HebelerAstroPostL & \HebelerAstroPostKsym & \HebelerAstroPostRskin & \HebelerAstroPostAlphaD \\
         & Astro+PREX-II Post.
            & \HebelerAstroPREXPostEpnm & \HebelerAstroPREXPostS & \HebelerAstroPREXPostL & \HebelerAstroPREXPostKsym & \HebelerAstroPREXPostRskin &\HebelerAstroPREXPostAlphaD \\
         & Astro+$\alpha_D^{^{208}\mathrm{Pb}}$ Post.
            & \HebelerAstroAlphaDPostEpnm & \HebelerAstroAlphaDPostS & \HebelerAstroAlphaDPostL & \HebelerAstroAlphaDPostKsym & \HebelerAstroAlphaDPostRskin &\HebelerAstroAlphaDPostAlphaD \\
        \cline{2-8}
        \multirow{5}{*}{\MBPT~\cite{Tews:2012fj}}
         & Original Work & \externalresult{[13.4, 20.1]} & \externalresult{[28.9, 34.9]} & \externalresult{[43.0, 66.6]} & - & -  & - \\
         & Prior
            & \mbptPriorEpnm & \mbptPriorS & \mbptPriorL & \mbptPriorKsym & \mbptPriorRskin & \mbptPriorAlphaD \\
         & Astro Posterior
            & \mbptAstroPostEpnm & \mbptAstroPostS & \mbptAstroPostL & \mbptAstroPostKsym & \mbptAstroPostRskin & \mbptAstroPostAlphaD \\
         & Astro+PREX-II Post.
            & \mbptAstroPREXPostEpnm & \mbptAstroPREXPostS & \mbptAstroPREXPostL & \mbptAstroPREXPostKsym & \mbptAstroPREXPostRskin & \mbptAstroPREXPostAlphaD \\
         & Astro+$\alpha_D^{^{208}\mathrm{Pb}}$ Post.
            & \mbptAstroAlphaDPostEpnm & \mbptAstroAlphaDPostS & \mbptAstroAlphaDPostL & \mbptAstroAlphaDPostKsym & \mbptAstroAlphaDPostRskin & \mbptAstroAlphaDPostAlphaD \\
        \cline{2-8}
        \multirow{5}{*}{\DrischlerPRL~\cite{Drischler:2017wtt}}
         & Original Work & \externalresult{[15.3, 18.7]} & - & - & - & - & - \\
         & Prior
            & \DrischlerPRLPriorEpnm & \DrischlerPRLPriorS & \DrischlerPRLPriorL & \DrischlerPRLPriorKsym & \DrischlerPRLPriorRskin & \DrischlerPRLPriorAlphaD \\
         & Astro Posterior
            & \DrischlerPRLAstroPostEpnm & \DrischlerPRLAstroPostS & \DrischlerPRLAstroPostL & \DrischlerPRLAstroPostKsym & \DrischlerPRLAstroPostRskin & \DrischlerPRLAstroPostAlphaD \\
         & Astro+PREX-II Post.
            & \DrischlerPRLAstroPREXPostEpnm & \DrischlerPRLAstroPREXPostS & \DrischlerPRLAstroPREXPostL & \DrischlerPRLAstroPREXPostKsym & \DrischlerPRLAstroPREXPostRskin & \DrischlerPRLAstroPREXPostAlphaD \\
         & Astro+$\alpha_D^{^{208}\mathrm{Pb}}$ Post.
            & \DrischlerPRLAstroAlphaDPostEpnm & \DrischlerPRLAstroAlphaDPostS & \DrischlerPRLAstroAlphaDPostL & \DrischlerPRLAstroAlphaDPostKsym & \DrischlerPRLAstroAlphaDPostRskin & \DrischlerPRLAstroAlphaDPostAlphaD \\
        \hline
        \hline
         & Prior
            & \DrischlerGPPriorEpnm & \DrischlerGPPriorS & \DrischlerGPPriorL & \DrischlerGPPriorKsym & \DrischlerGPPriorRskin & \DrischlerGPPriorAlphaD \\
        \DrischlerGP & Astro Posterior
            & \DrischlerGPAstroPostEpnm & \DrischlerGPAstroPostS & \DrischlerGPAstroPostL & \DrischlerGPAstroPostKsym & \DrischlerGPAstroPostRskin & \DrischlerGPAstroPostAlphaD \\
        \cite{Drischler:2020, Drischler:2020prc} & Astro+PREX-II Post.
            & \DrischlerGPAstroPREXPostEpnm & \DrischlerGPAstroPREXPostS & \DrischlerGPAstroPREXPostL & \DrischlerGPAstroPREXPostKsym & \DrischlerGPAstroPREXPostRskin & \DrischlerGPAstroPREXPostAlphaD \\
        & Astro+$\alpha_D^{^{208}\mathrm{Pb}}$ Post.
            & \DrischlerGPAstroAlphaDPostEpnm & \DrischlerGPAstroAlphaDPostS & \DrischlerGPAstroAlphaDPostL & \DrischlerGPAstroAlphaDPostKsym & \DrischlerGPAstroAlphaDPostRskin & \DrischlerGPAstroAlphaDPostAlphaD \\
        \hline
    \end{tabular}
    }
    \caption{
        Medians and 90\% highest-probability-density credible regions for selected nuclear properties.
        All $\EFT$ results trust the theoretical prediction up to $p_\mathrm{max}/c^2=4.3\times10^{12}\,\mathrm{g}/\mathrm{cm}^3$, corresponding to $n(p_\mathrm{max})\sim n_0$.
        \EFT-marginalized results combine results from \QMC~\cite{Lynn:2016}, \Hebeler~\cite{Hebeler:2009iv,Hebeler:2013nza}, \MBPT~\cite{Tews:2012fj}, and \DrischlerPRL~\cite{Drischler:2017wtt} with equal weight \textit{a priori}.
        We also tabulate results from each of these 4 \EFT~predictions separately.
        In addition, we provide results from \DrischlerGP~\cite{Drischler:2020, Drischler:2020prc} for comparison with \DrischlerPRL, both of which use the same microscopic calculations.
        Where possible, we also provide bounds quoted for the original studies, given by envelopes containing all models considered within the original studies.
        As such, they do not have an immediate statistical interpretation and are wider than our 90\% credible regions.
    }
    \label{tab:credible regions}
\end{table*}

We first summarize our conclusions about $R_\mathrm{skin}^{^{208}\mathrm{Pb}}$ in Sec.~\ref{sec:astro results} before comparing constraints on broader sets of nuclear properties near $n_0$ in Sec.~\ref{sec:compatibility for nuclear parameters}.
Section~\ref{sec:comparison for NS observables} summarizes what we can learn about NS properties from current experimental constraints and possible future improvements.

\subsection{Symmetry-Energy Parameters and Neutron-Skin Thickness in Lead}
\label{sec:astro results}

We begin by discussing our findings for $S_0$, $L$, $K_{\rm sym}$, and $R_\mathrm{skin}^{^{208}{\rm Pb}}$, shown in Fig.~\ref{fig:decomposed esym corner}.
We plot the nonparametric prior, the posterior constrained only by astrophysical data, and the posterior additionally constrained by $\chi$EFT calculations up to $n\approx n_0$. 
Our GPs are conditioned on $\chi$EFT up to a maximum pressure, $p_\mathrm{max}$. 
To translate this into a density, we report the median density at $p_\mathrm{max}$ \textit{a priori}; the exact density at $p_\mathrm{max}$ varies due to uncertainty in the EOS from $\chi$EFT.
In addition to the constraints obtained by marginalizing over the four separate \EFT~calculations, we also show the posteriors for each individual $\chi$EFT calculation.
Finally, we also compare our results with the recent constraints on $R_\mathrm{skin}^{^{208}{\rm Pb}}$ and $L$ from the PREX-II experiment~\cite{PREXII}, where we have translated from $R_\mathrm{skin}^{^{208}{\rm Pb}}$ to $L$ using our model of the theoretical uncertainty in the correlation between these two quantities.
Prior and posterior credible regions are also provided in Table~\ref{tab:credible regions}.

The priors and Astro-only posteriors for the nonparametric inference are very broad, and we find large ranges for $S_0$, $L$, $K_{\rm sym}$, and $R_\mathrm{skin}^{^{208}{\rm Pb}}$ (see Table~\ref{tab:credible regions}).
The astrophysical data slightly informs our uncertainty in $S_0$ and $L$, shifting the median values of their distributions, but the 90\% confidence intervals are less impacted.
The astrophysical data does not strongly constrain $K_{\rm sym}$, but suggests that it is negative.
Taken together, this highlights the fact that astrophysical information alone is not sufficient to pin down properties of the EOS around nuclear saturation density. 

When we additionally constrain the nonparametric EOSs using the four $\chi$EFT calculations, we obtain much narrower posteriors.
It is noteworthy that the \EFT~posteriors fall near the maximum of the Astro-only nonparametric posterior.
We stress that this need not have been the case, because the nonparametric Astro-only posterior does not know anything about \EFT.
While the four individual calculations result in slightly different values for $L$ and, hence, $R_\mathrm{skin}^{^{208}{\rm Pb}}$, overall all four $\chi$EFT calculations are very consistent.

When we compare our findings with the recent PREX-II results, we find that the nonparametric Astro-only posterior prefers lower values for $L$ and $R_\mathrm{skin}^{^{208}{\rm Pb}}$, in good agreement with the result that includes \EFT.
Both posteriors peak at similar values of $L$, on the order of $50$--$60$~MeV, and of $R_\mathrm{skin}^{^{208}{\rm Pb}}$, on the order of $0.15$--$0.20$~fm.
However, uncertainties are large and nonparametric Astro-only results remain compatible with both the $\chi$EFT prediction and the PREX-II results.
Nonetheless, when we additionally condition on $\chi$EFT calculations, we find that the PREX-II result for $R_\mathrm{skin}^{^{208}{\rm Pb}}$ and the associated range for $L$ ($69$--$143$~MeV at $1\sigma$~\cite{Reed:2021nqk}), are only in mild tension with the \EFT~predictions.

Finally, we compare our findings for $S_0$ and $L$ with other constraints in the upper-right panel of Fig.~\ref{fig:decomposed esym corner}.
Our \EFT+Astro posterior is very consistent with the overlap region from various experimental constraints from Lattimer and Prakash~\cite{LATTIMER2016127} and lies fully witin the bounds of the unitary gas conjecture~\cite{Tews:2016jhi}.
While the extraction of $S_0$ and $L$ from PREX-II by Reed \textit{et al.}~\cite{Reed:2021nqk} leads to significantly larger central values, it also has large 90\% credible regions, which overlap with our \EFT+Astro posterior.
In addition, we show here the correlation obtained from the experimental value of the dipole polarizability $\alpha_D^{^{208}\mathrm{Pb}}$~\cite{Tamii:2011pv} with our uncertainty model Eq.~\eqref{eq:L2alphDS0} assuming uninformative priors for $S_0$ and $L$.
This overlaps nicely with all extractions.

\subsection{Compatibility of Astrophysical, Experimental, and Theoretical Results for Nuclear Properties}
\label{sec:compatibility for nuclear parameters}

\begin{figure}
    \includegraphics[width=1.0\columnwidth, clip=True, trim=0.6cm 1.55cm 0.5cm 0.0cm]{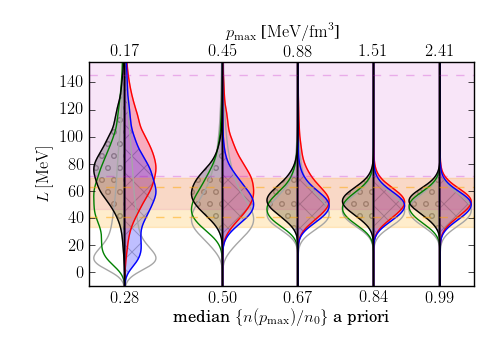} \\
    \includegraphics[width=1.0\columnwidth, clip=True, trim=0.6cm 1.55cm 0.5cm 1.5cm]{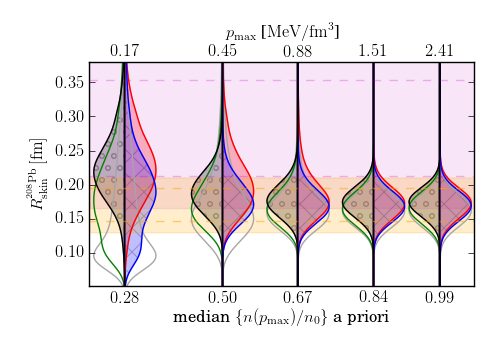} \\
    \includegraphics[width=1.0\columnwidth, clip=True, trim=0.6cm 0.0cm 0.5cm 1.5cm]{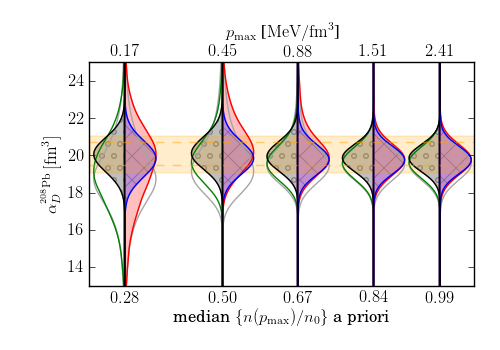}
    \caption{
        Priors (\emph{grey, unshaded}), Astro-only posteriors (\emph{left side of violins, green unshaded}), Astro+PREX-II posteriors (\emph{right side of violins, red shaded}), Astro+$\alpha_D^{^{208}\mathrm{Pb}}$ posteriors (\emph{right side of violins, blue shaded+hatched}), and Astro+PREX-II+$\alpha_D^{^{208}\mathrm{Pb}}$ posteriors (\emph{left side of violins, grey shaded+dots}) for \EFT-marginalized results as a function of the maximum pressure up to which we trust \EFT.
        The left-most curves (median $n\sim 0.3 n_0$) are equivalent to the nonparametric results in Fig.~\ref{fig:decomposed esym corner}.
        Horizontal bands (dashed lines) correspond to 90\% (1-$\sigma$) credible regions from PREX-II~\cite{PREXII} ($R_\mathrm{skin}^{^{208}\mathrm{Pb}}$; \emph{pink}) and the electric dipole polarizability~\cite{Tamii:2011pv} ($\alpha_D^{^{208}\mathrm{Pb}}$; \emph{orange}).
        When translating experimental data to their correlated properties in this figure (e.g., horizontal $\alpha_D^{^{208}\mathrm{Pb}}$ bands for $L$ and $R_\mathrm{skin}^{^{208}{\rm Pb}}$), we employ our uncertainty relations in the theoretical correlations (Eqs.~\eqref{eq:L2Rskin} and~\eqref{eq:L2alphDS0}, assuming $S_0=32.5\,\mathrm{MeV}$ for the latter).
    }
    \label{fig:violins}
\end{figure}

In Fig.~\ref{fig:violins}, we show the evolution of our constraints on $L$, $R_\mathrm{skin}^{^{208}{\rm Pb}}$, and $\alpha_D^{^{208}\mathrm{Pb}}$ as a function of the maximum density up to which we condition our prior on $\chi$EFT. 
In addition to the posterior conditioned only on astrophysical data, we show results for three cases that are additionally conditioned on either the PREX-II $R_\mathrm{skin}^{^{208}\mathrm{Pb}}$ data~\cite{PREXII}, the $\alpha_D^{^{208}\mathrm{Pb}}$ data from Ref.~\cite{Tamii:2011pv}, or both.

\begin{figure}
    \centering
    \includegraphics[width=1.0\columnwidth, clip=True, trim=0.0cm 0.00cm 0.0cm 0.30cm]{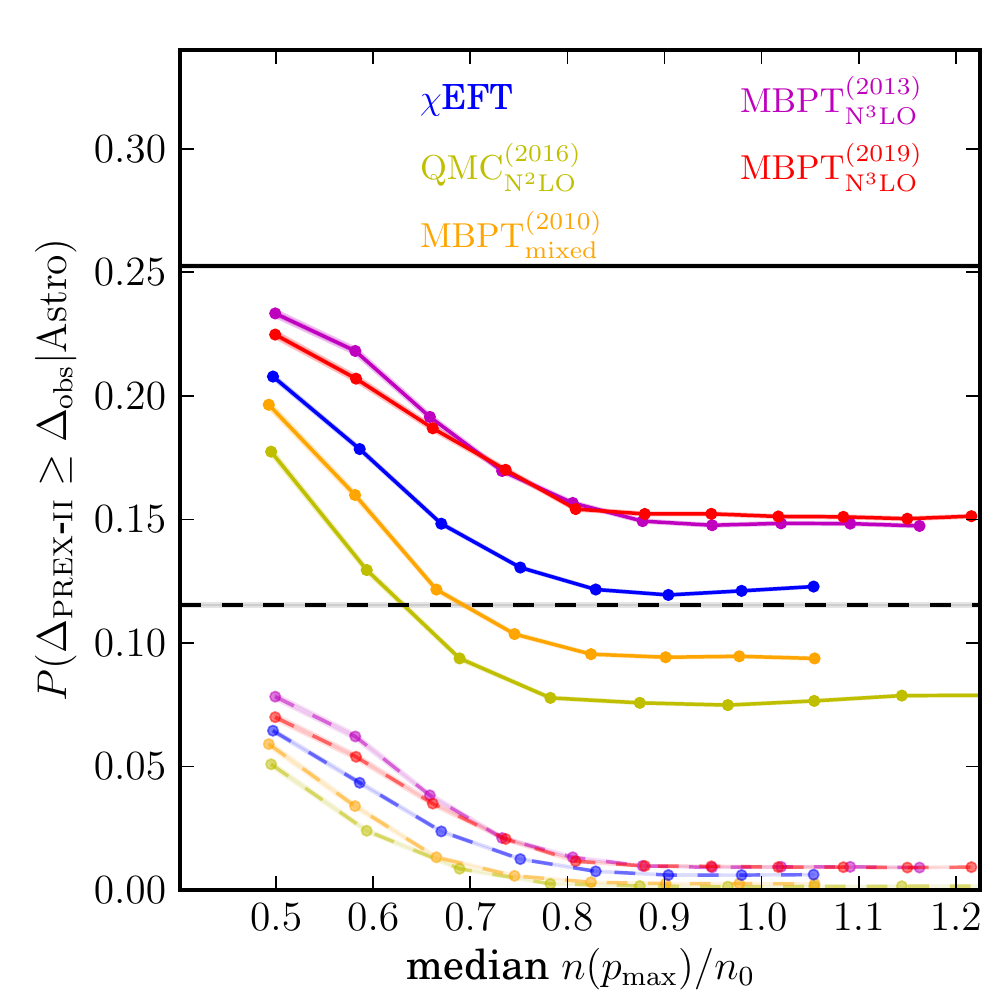}
    \caption{
        Probability of PREX-II disagreeing with posteriors conditioned on \EFT~up to $p_\mathrm{max}$ by at least the measured difference given experimental uncertainties ($p$-values, \emph{solid lines}).
        We also show the $p$-values for a hypothetical experiment producing the same mean as PREX-II with half the uncertainty (\emph{dashed lines}).
        Results are given for nonparametric Astro-only posteriors (\emph{black horizontal lines}), \EFT-marginalized (\emph{blue}), \QMC~(\emph{yellow}), \Hebeler~(\emph{orange}), \MBPT~(\emph{purple}), and \DrischlerPRL~(\emph{red}).
    }
    \label{fig:pvalue}
\end{figure}

\begin{figure}
    \includegraphics[width=1.0\columnwidth, clip=True, trim=0.0cm 1.3cm 0.0cm 0.2cm]{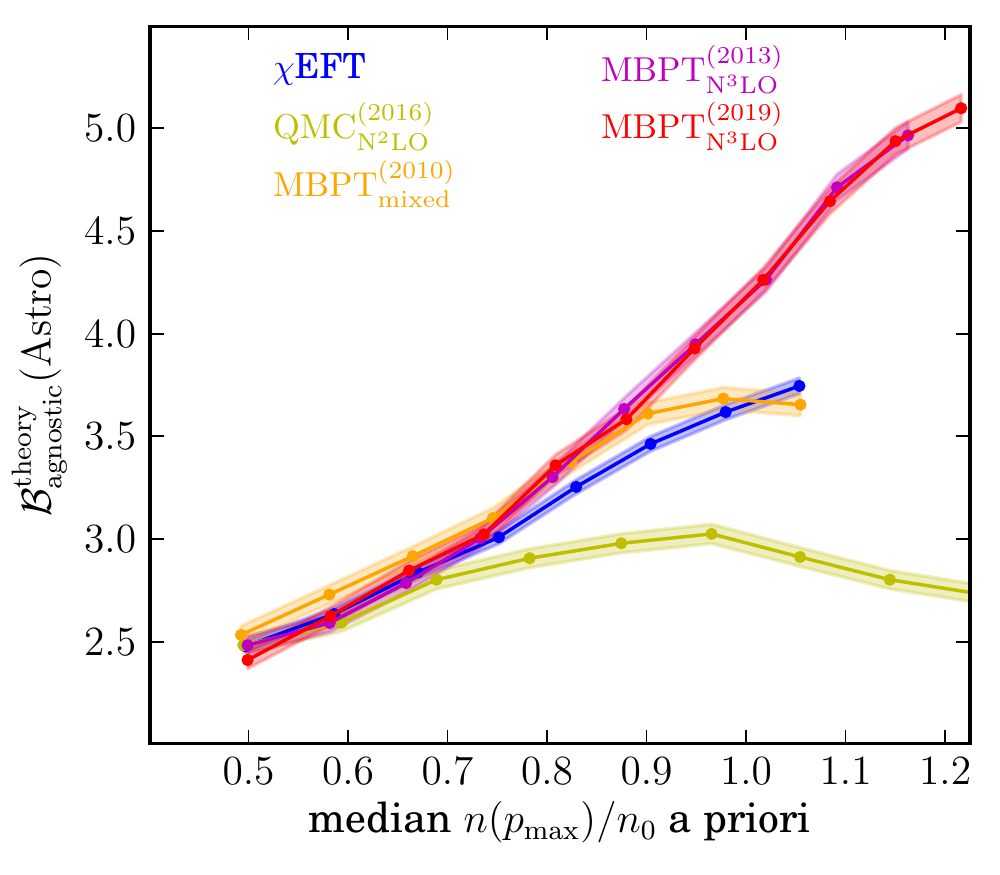} \\
    \includegraphics[width=1.0\columnwidth, clip=True, trim=0.0cm 1.3cm 0.0cm 0.2cm]{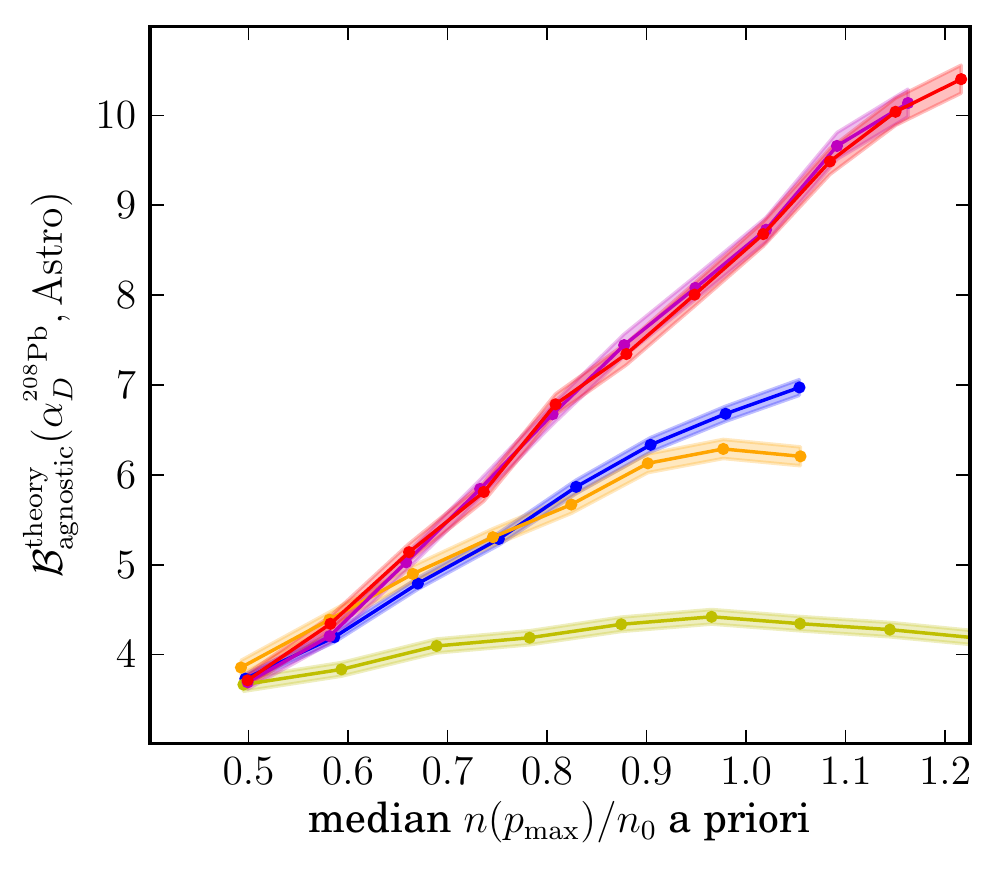} \\
    \includegraphics[width=1.0\columnwidth, clip=True, trim=0.0cm 0.0cm 0.0cm 0.2cm]{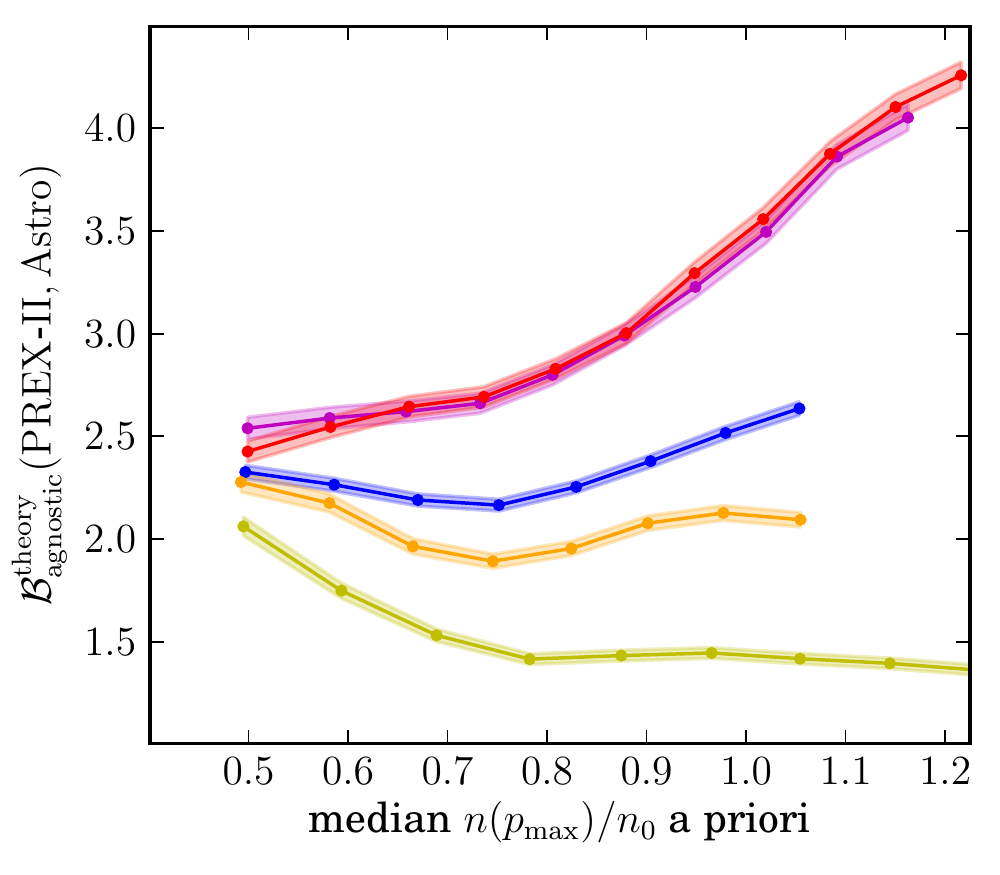}
    \caption{
        Bayes factors between priors conditioned on \EFT~calculations up to different $p_\mathrm{max}$ vs. the priors not conditioned on \EFT~at all for (\emph{top}) astrophysical data, (\emph{middle}) Astro+$\alpha_D^{^{208}\mathrm{Pb}}$, and (\emph{bottom}) Astro+PREX-II data.
        We show results for the \EFT-marginalized calculations (\emph{blue}) as well as the  \QMC~(\emph{yellow}), \Hebeler~(\emph{orange}), \MBPT~(\emph{purple}), and  \DrischlerPRL~(\emph{red}) calculations separately.
    }
    \label{fig:evidence}
\end{figure}

\begin{figure}
    \includegraphics[width=1.0\columnwidth, clip=True, trim=0.0cm 1.3cm 0.0cm 0.2cm]{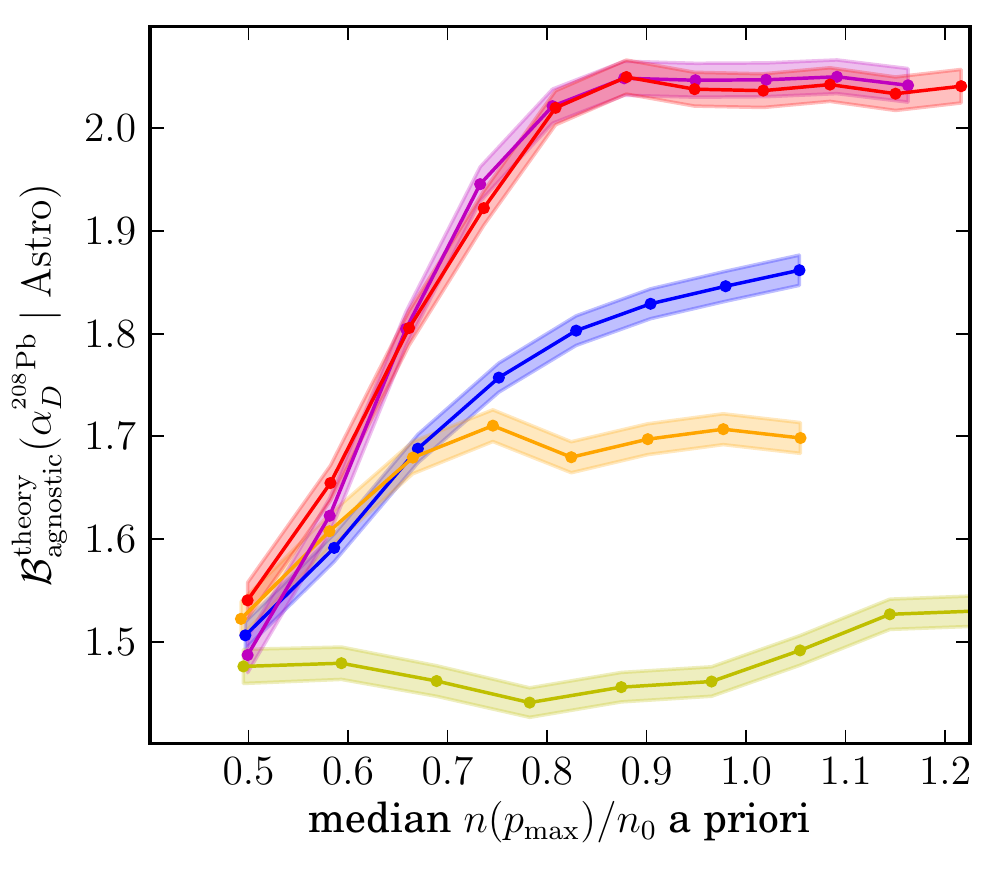}
    \includegraphics[width=1.0\columnwidth, clip=True, trim=0.0cm 0.0cm 0.0cm 0.2cm]{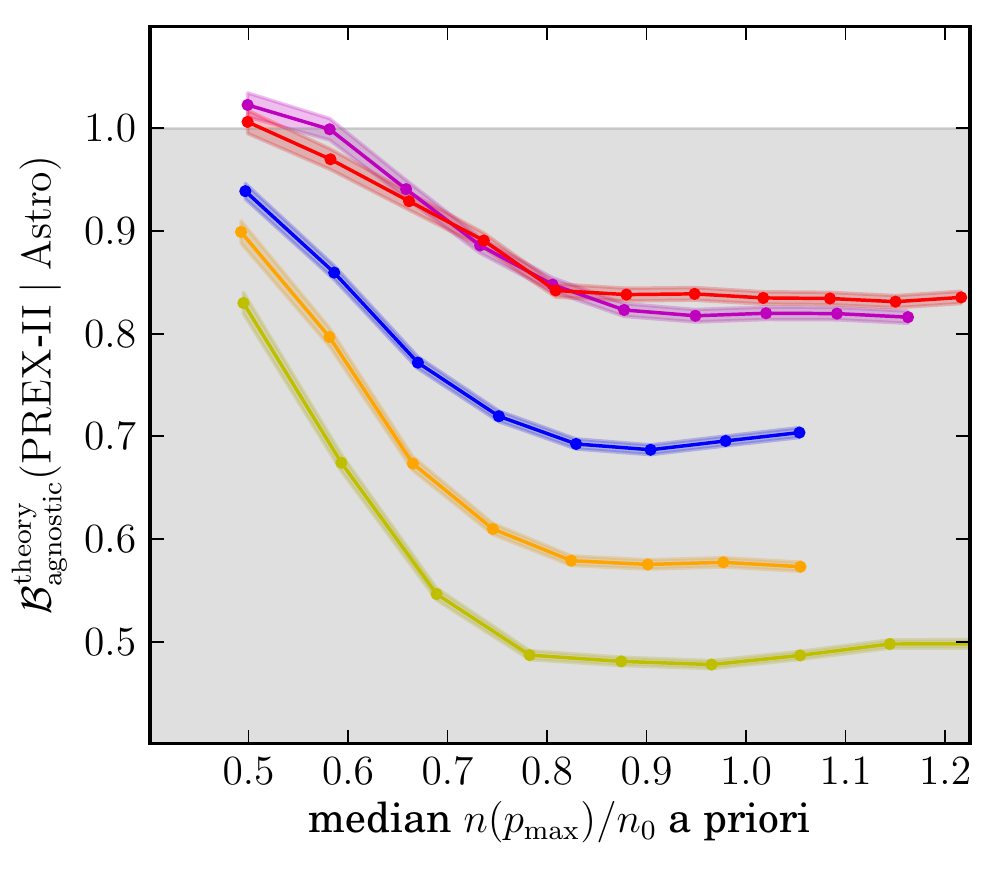} \\
    \vspace{-7.05cm}
    \hspace{4.0cm} \textbf{Theory Favored} \\
    \vspace{0.25cm}
    \hspace{4.0cm} \textbf{Theory Disfavored} \\
    \vspace{+6.0cm}
    \caption{
        Bayes factors between priors conditioned on \EFT~vs. priors not conditioned on \EFT~at all for different nuclear data when we first condition on the astrophysical observations (include them as part of the prior).
        We show the result for (\emph{top}) $\alpha_D^{^{208}\mathrm{Pb}}$ and (\emph{bottom}) PREX-II data.
    }
    \label{fig:conditioned evidence}
\end{figure}

\begin{figure}
    \centering
    \includegraphics[width=0.9\columnwidth, clip=True, trim=0.25cm 2.00cm 0.50cm 0.00cm]{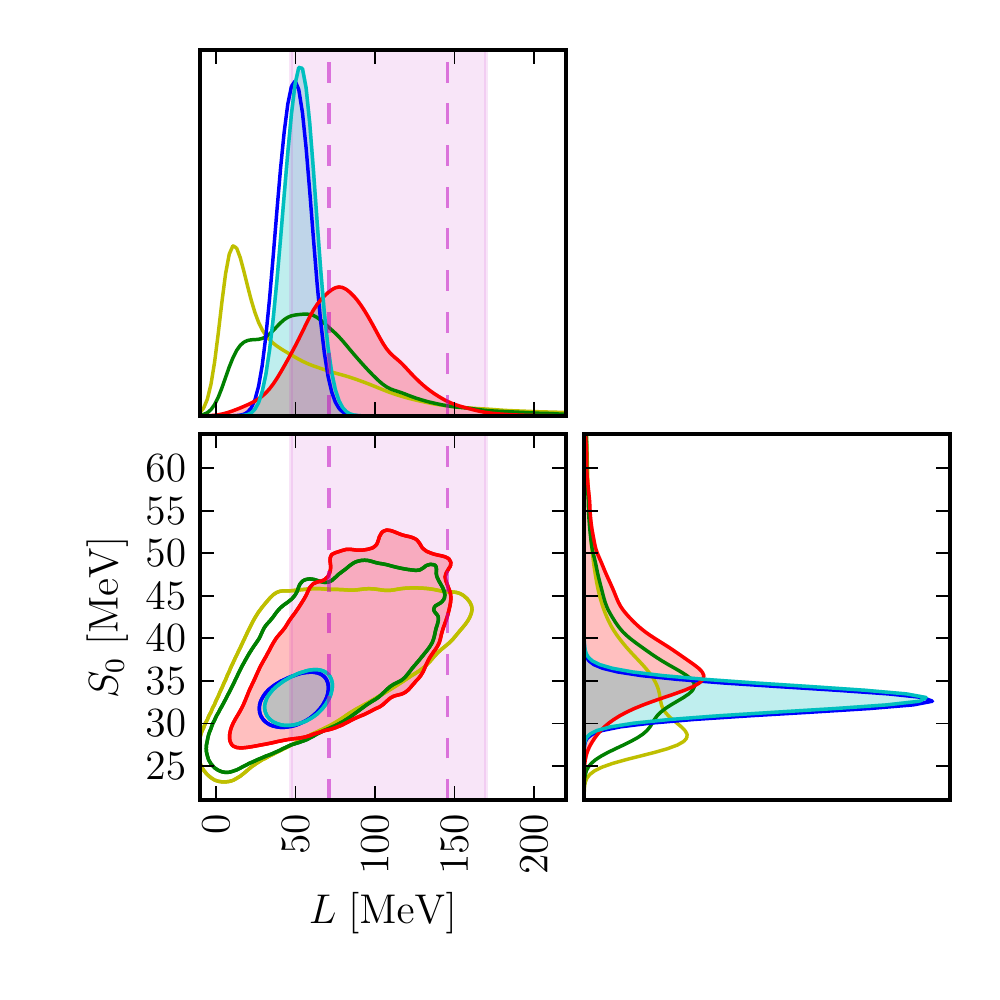} \\
    \includegraphics[width=0.9\columnwidth, clip=True, trim=0.25cm 2.00cm 0.50cm 4.30cm]{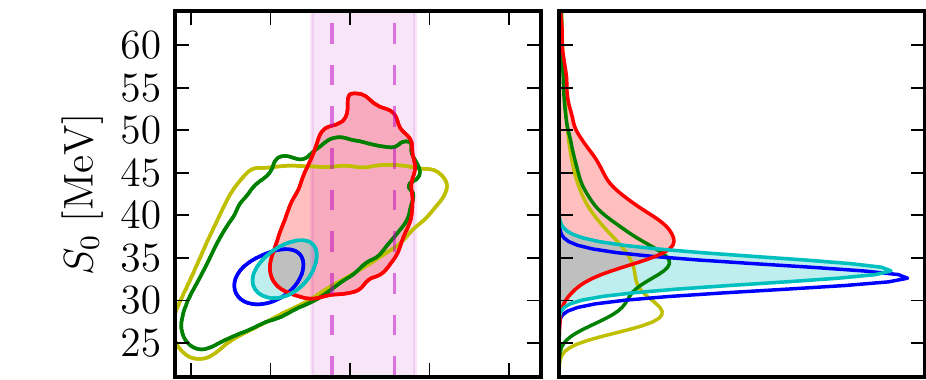} \\
    \includegraphics[width=0.9\columnwidth, clip=True, trim=0.25cm 0.50cm 0.50cm 4.30cm]{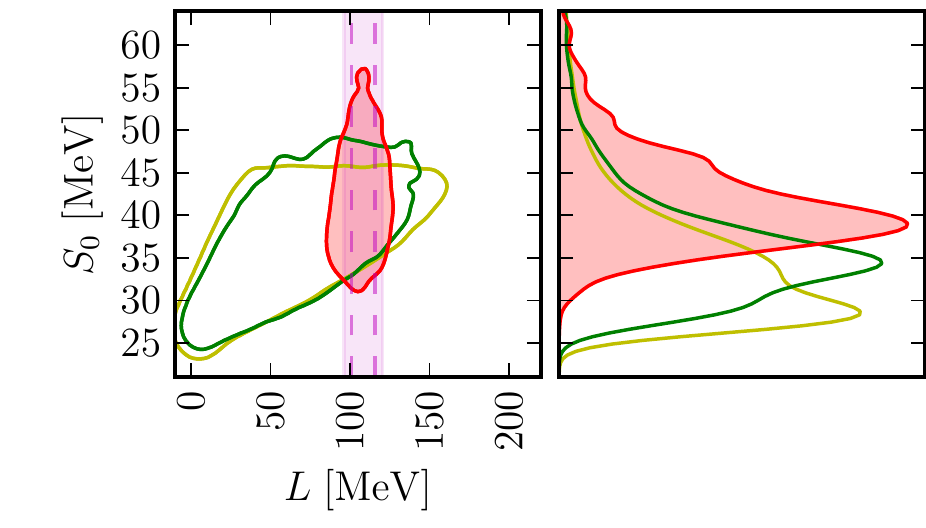} \\
    \vspace{-14.00cm}
    \hfill \textcolor{red}{Nonparametric} \\
    \hfill \textcolor{red}{Astro+PREX-II} \\
    \hfill \textcolor{red}{Posterior} \\
    \vspace{+0.25cm}
    \hfill \textcolor{cyan}{\EFT} \\
    \hfill \textcolor{cyan}{Astro+PREX-II} \\
    \hfill \textcolor{cyan}{Posterior} \\
    \vspace{+11.25cm}
    \caption{
        Correlations between $S_0$ and $L$ when we model the PREX-II estimate with different uncertainties: (\emph{top}) the actual measurement uncertainty, (\emph{middle}) a hypothetical measurement with half the PREX-II uncertainty, and (\emph{bottom}) a hypothetical measurement with vanishingly small uncertainty for $R_\mathrm{skin}^{^{208}\mathrm{Pb}}$.
        We show the nonparametric prior (\emph{unshaded yellow}), Astro-only posterior (\emph{unshaded green}), and Astro+PREX-II posterior (\emph{shaded red}) as well as the \EFT-marginalized Astro-only posterior (\emph{unshaded blue}) and Astro+PREX-II posterior (\emph{shaded light blue}).
        As in Fig.~\ref{fig:decomposed esym corner}, (\emph{pink}) shaded vertical bands represent (real and hypothetical) PREX-II 90\% credible regions and dashed lines show the 1-$\sigma$ credible regions uncertainty.
        Improved measurements of $R_\mathrm{skin}^{^{208}\mathrm{Pb}}$ are still consistent with a wide range of $S_0$ within the nonparametric inference.
    }
    \label{fig:S0 and L}
\end{figure}

If we do not condition the prior on \EFT~(left-most violins, where we match directly to the crust at $0.3 n_0$), the Astro-only posterior retains large uncertainties for all three quantities. 
As stated before, astrophysical data inform our knowledge of $L$ and $R_\mathrm{skin}^{^{208}{\rm Pb}}$ to some degree, but they do not add further information about $\alpha_D^{^{208}\mathrm{Pb}}$ because $S_0$ is not strongly constrained.
When we additionally condition on the recent PREX-II result, uncertainties remain large, but the posteriors for $L$ and $R_\mathrm{skin}^{^{208}{\rm Pb}}$ are pushed to higher values. 
Alternatively, conditioning instead on the $\alpha_D^{^{208}\mathrm{Pb}}$ measurement, the posteriors for $L$ and $R_\mathrm{skin}^{^{208}{\rm Pb}}$ agree very well with the Astro-only result, highlighting the consistency of this experiment and neutron-star observations; see also Table~\ref{tab:credible regions}.
In this case, as expected, the posterior for $\alpha_D^{^{208}\mathrm{Pb}}$ is much narrower.
Conditioning on astrophysical observations and both PREX-II and $\alpha_D^{^{208}\mathrm{Pb}}$ produces posteriors for $L$ and $R_\mathrm{skin}^{^{208}{\rm Pb}}$ similar to those obtained by only conditioning on astrophysical observations and PREX-II because there is enough additional freedom in $S_0$ to accomodate the $\alpha_D^{^{208}\mathrm{Pb}}$ measurements for almost any $L$ (see also Fig.~\ref{fig:S0 and L}).

When conditioning the priors on $\chi$EFT constraints to higher densities, all posteriors start to overlap more.
They agree with each other very closely if we condition up to $n_0$, where the \EFT~constraints dominate. 
In this case, the tension of our process with the PREX-II results is maximized but nonetheless remains mild due to the large PREX-II uncertainties.
On the other hand, the agreement with the $\alpha_D^{^{208}\mathrm{Pb}}$ result improves the more we trust the $\chi$EFT constraints.

\begin{figure*}
    \begin{minipage}{0.366\textwidth}
        \begin{center}
            \large{Prior} \\
            \includegraphics[width=1.0\textwidth, clip=True, trim=0.0cm 1.45cm 0.40cm 0.20cm]{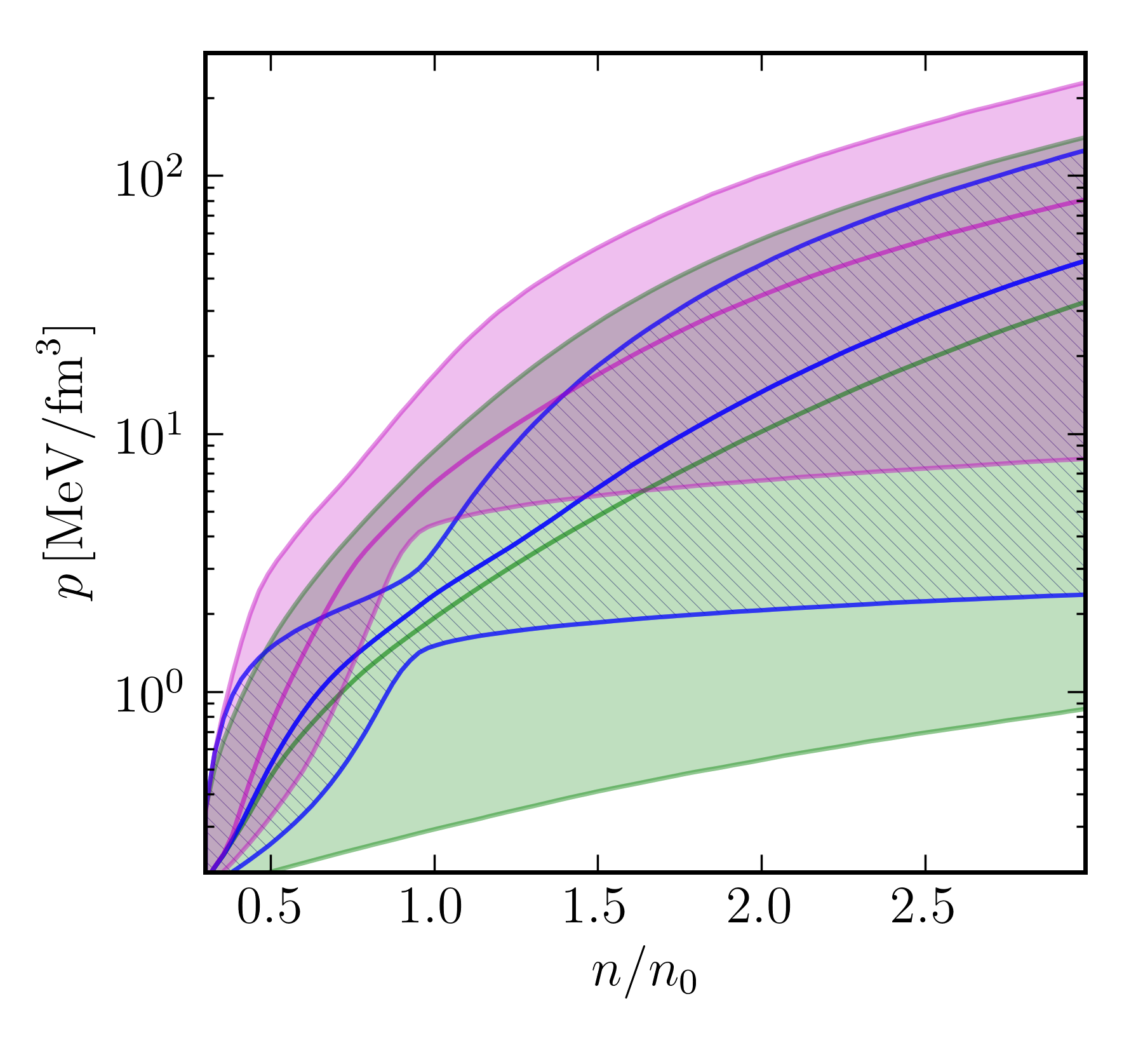}
            \includegraphics[width=1.0\textwidth, clip=True, trim=0.0cm 0.00cm 0.40cm 0.20cm]{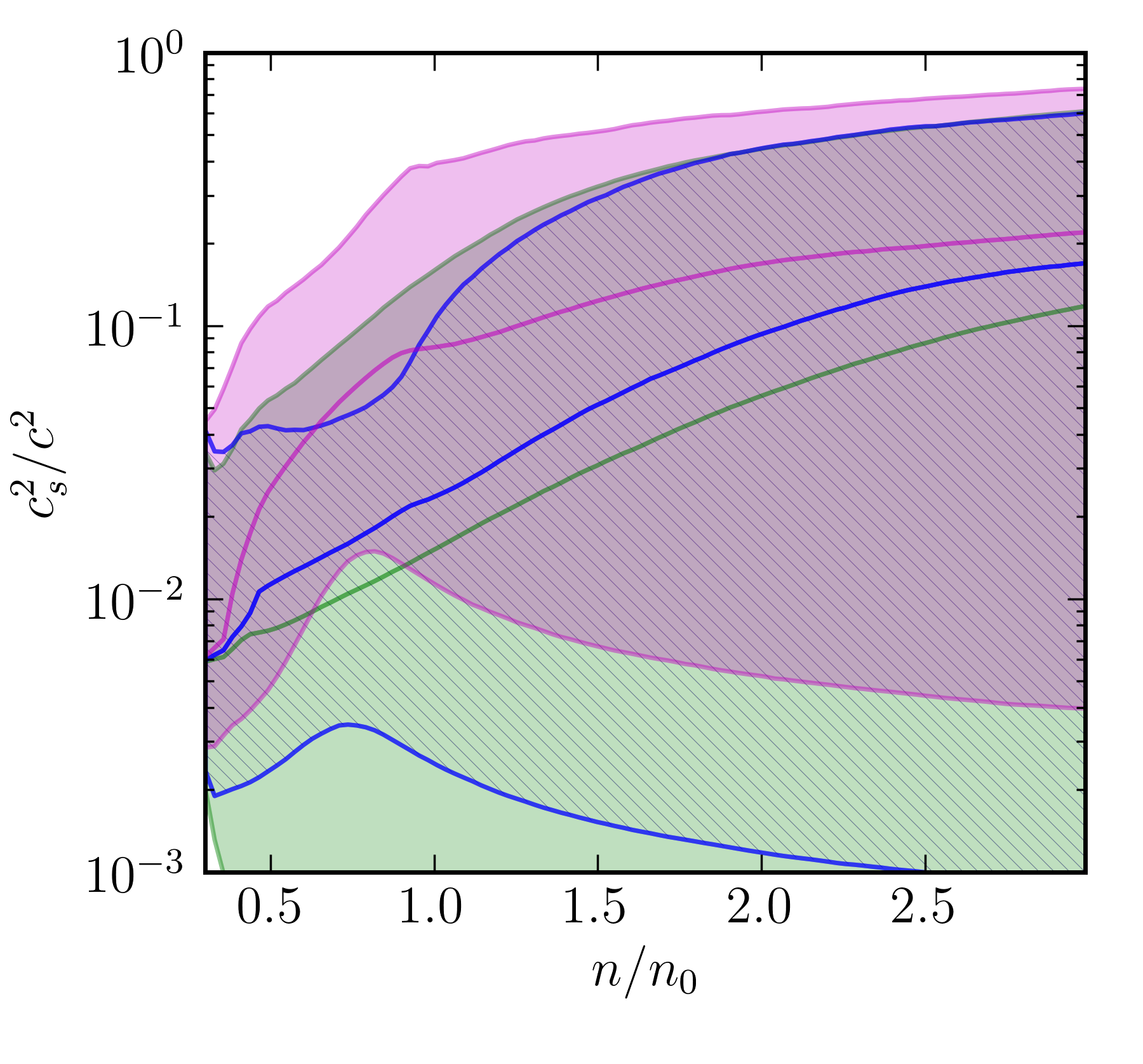}
        \end{center}
    \end{minipage}
    \begin{minipage}{0.305\textwidth}
        \begin{center}
            \large{Astro Posterior} \\
            \includegraphics[width=1.0\textwidth, clip=True, trim=1.50cm 1.45cm 0.40cm 0.20cm]{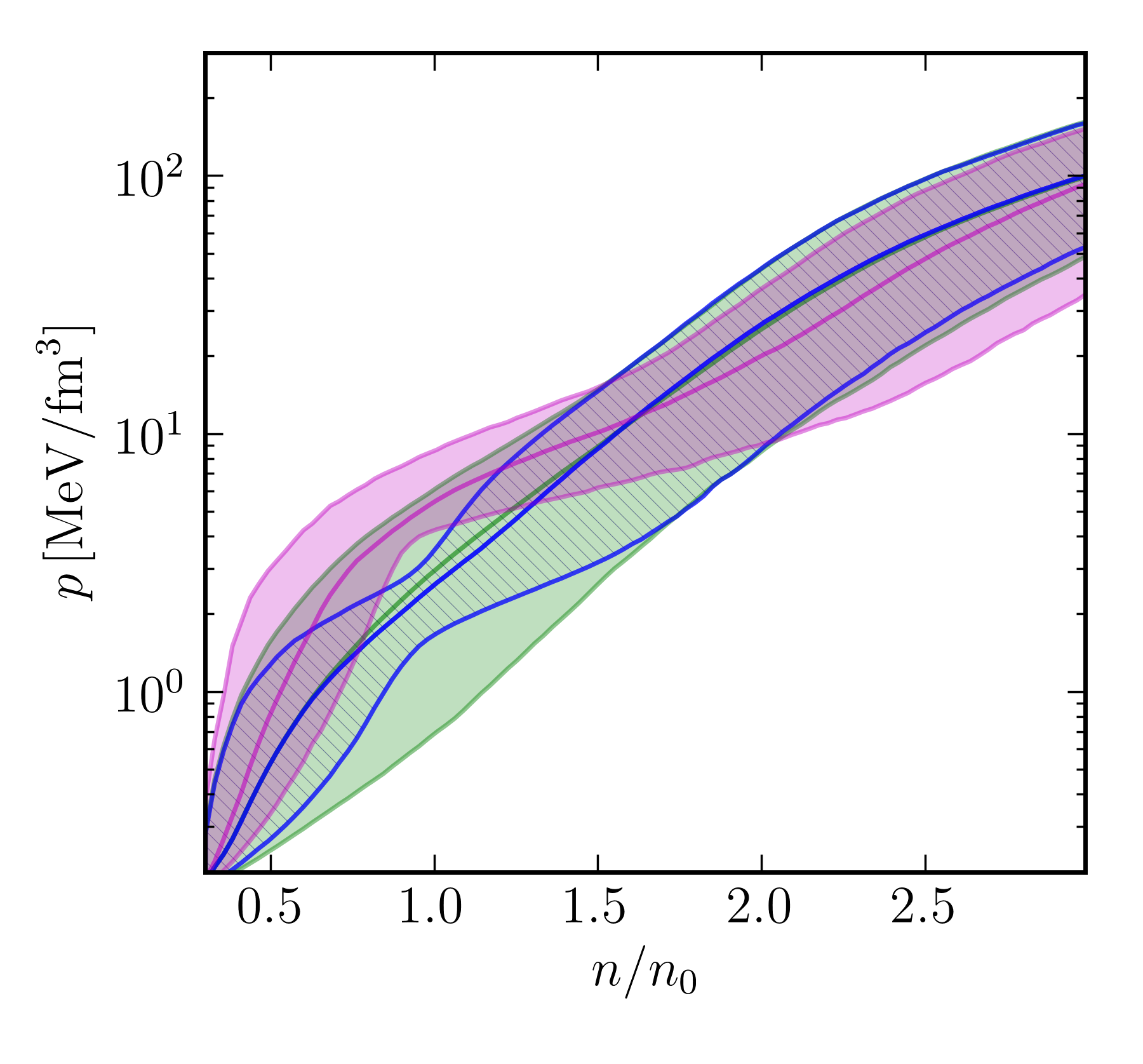}
            \includegraphics[width=1.0\textwidth, clip=True, trim=1.50cm 0.00cm 0.40cm 0.20cm]{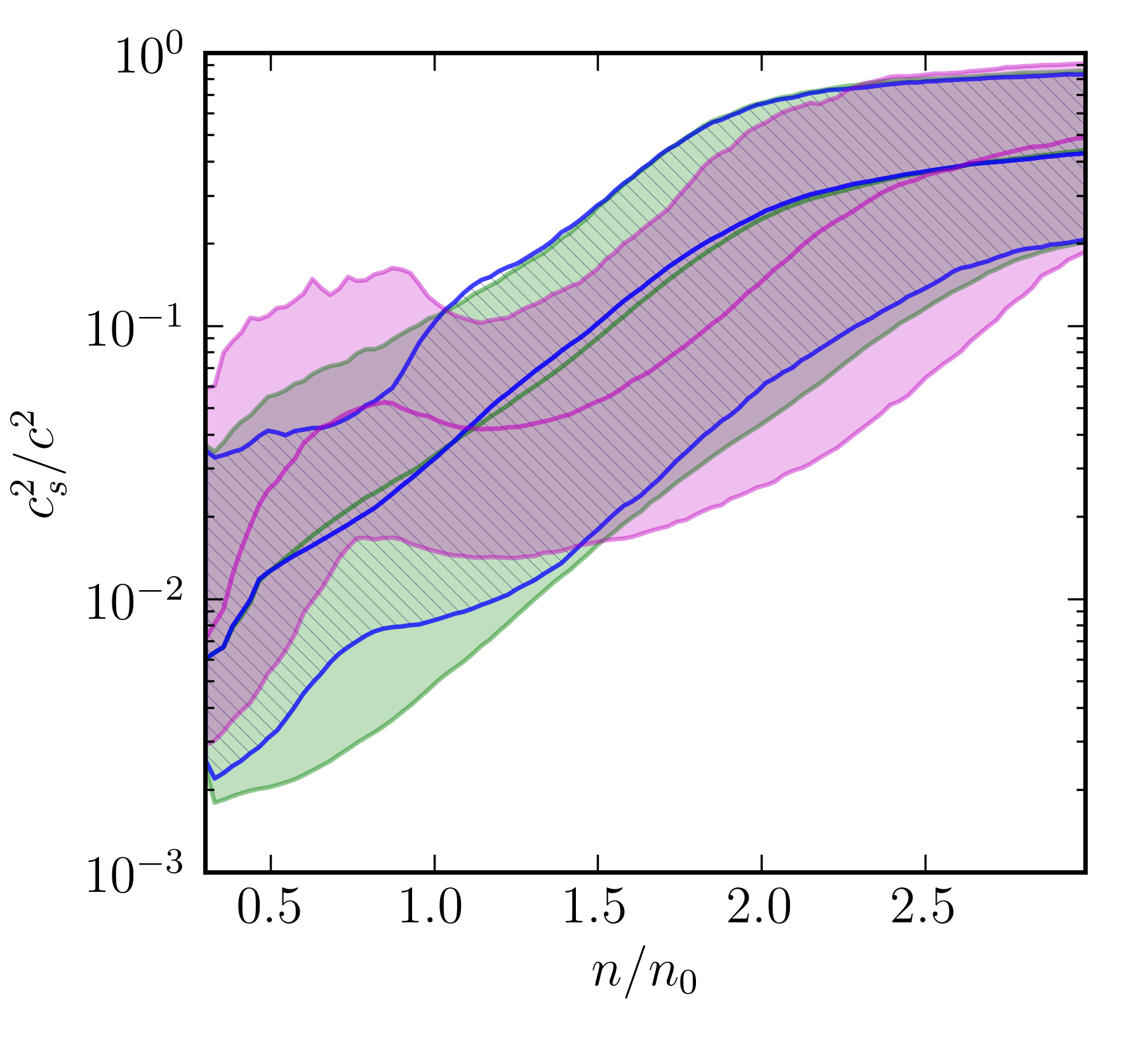}
        \end{center}
    \end{minipage}
    \begin{minipage}{0.305\textwidth}
        \begin{center}
            \large{Astro+PREX-II Posterior} \\
            \includegraphics[width=1.0\textwidth, clip=True, trim=1.50cm 1.45cm 0.40cm 0.20cm]{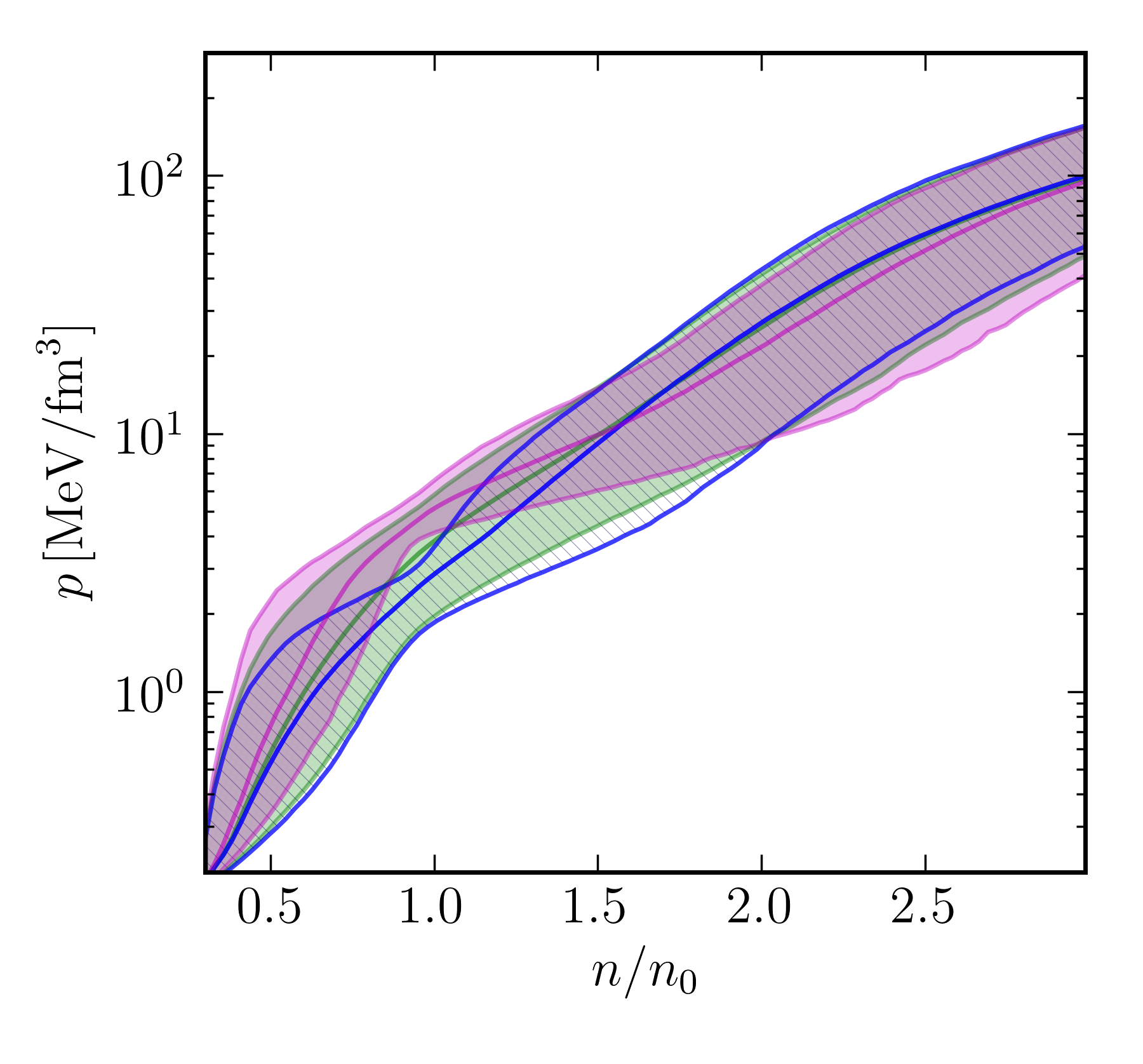}
            \includegraphics[width=1.0\textwidth, clip=True, trim=1.50cm 0.00cm 0.40cm 0.20cm]{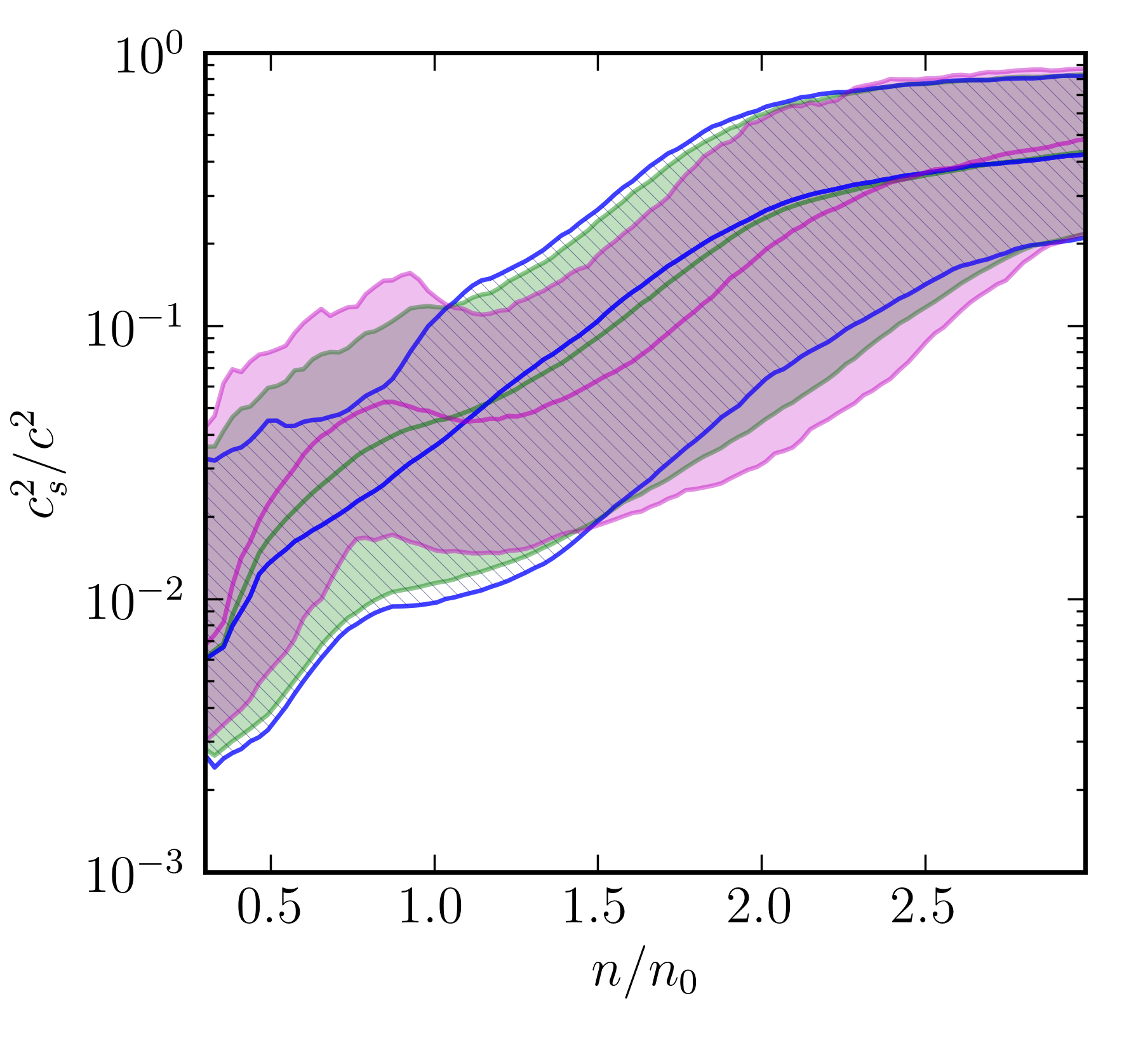}
        \end{center}
    \end{minipage}
    \caption{
        Median and 90\% symmetric credible regions for the prior (\emph{left}), Astro-only posterior (\emph{middle}), and Astro+PREX-II posterior (\emph{right}) for all EOS and all values of $L$ (\emph{green}), EOS with $30\,\mev < L < 70\,\mev$ (\emph{hatched blue}), and EOS with $100\,\mev < L$ (\emph{purple}).
        The main effect of the PREX-II data is to rule out some of the very soft EOS at low densities ($L \lesssim 30\,\mev$).
    }
    \label{fig:pressure-density with L cuts}
\end{figure*}

Figure~\ref{fig:pvalue} shows how the probability ($p$-value) that the true $R_\mathrm{skin}^{^{208}{\rm Pb}}$ differs from the PREX-II mean at least as much as the Astro+\EFT~posterior suggests, given the uncertainty in PREX-II's measurement.
The $p$-values decrease as we trust \EFT~up to higher densities, and we estimate a p-value of \mrgeftAstroPostPREXpvalue~when trusting \EFT~up to $n\sim n_0$ (c.f., \mrgagnAstroPostPREXpvalue~for the nonparametric Astro-only posterior).
However, if a hypothetical experiment confirmed the PREX-II mean value with half the uncertainty, this $p$-value would be reduced to \mrgeftAstroPostPREXhalfpvalue.
In fact, a hypothetical $R_\mathrm{skin}^{^{208}\mathrm{Pb}}$ measurement with half the uncertainty has a smaller $p$-value under the nonparametric Astro-only posterior than the \EFT-marginalized posterior has with the current $R_\mathrm{skin}^{^{208}\mathrm{Pb}}$ measurement uncertainties.

To investigate this further, we compute Bayes factors between the processes conditioned on \EFT~up to various pressures vs.~processes not conditioned on $\chi$EFT at all (Figs.~\ref{fig:evidence} and~\ref{fig:conditioned evidence}) for different sets of data: Astro-only, Astro+$\alpha_D^{^{208}\mathrm{Pb}}$, and Astro+PREX-II (Fig.~\ref{fig:evidence}) and when astrophysical data is already included in the prior (Fig.~\ref{fig:conditioned evidence}). 
In addition to the posteriors marginalized over all four $\chi$EFT results, we also show the Bayes factors for the individual \QMC,  \Hebeler, \MBPT, and \DrischlerPRL~results.
These Bayes factors quantify the relative likelihood of obtaining the observed data under different models, specifically whether \EFT=informed priors are more ($\mathcal{B}^\mathrm{theory}_\mathrm{agnostic}>1$) or less ($\mathcal{B}^\mathrm{theory}_\mathrm{agnostic}<1$) likely to have produced the observed data compared to our completely nonparametric prior.

Considering only astrophysical data, we find that $\chi$EFT is preferred over the theory-agnostic result up to at least nuclear saturation density. 
This is also true for the individual calculations, although we find that the Bayes factor in favor of \MBPT~and \DrischlerPRL~are a factor of two larger than for \QMC. 
This agrees with previous results~\cite{EssickTewsLandryReddyHolz2020} and could be associated with the higher-order \EFT~interactions included in \MBPT~and \DrischlerPRL~that tend to increase the pressure and are not included in \QMC.
It could also be associated with the different regularization schemes employed in these calculations.
However, this preference may be due to different widths of the theoretical uncertainty bands within different \EFT~calculations.
These Bayes factors are likely driven partly by Occam factors where a wider prior is penalized even though all models may achieve similar maximum likelihoods.
For example, \EFT~yields a narrower prior which penalizes the freedom in the nonparametric model without \EFT.
Similarly, the \MBPT~and \DrischlerPRL~priors predict higher median pressures with smaller uncertainties than \QMC, and both effects will tend to increase the relative Bayes factor.
We also find that the astrophysical observations can only distinguish between individual \EFT~calculations if we trust them up to $\gtrsim 0.75 n_0$.

When additionally including $\alpha_D^{^{208}\mathrm{Pb}}$, the Bayes factors in favor of \EFT~increase by a factor of two. 
In contrast, including the PREX-II information decreases the Bayes factors by a factor of $\lesssim 2$.
Figure~\ref{fig:conditioned evidence} shows this behavior explicitly by first conditioning on the astrophysical observations, thereby isolating the new information obtained from the inclusion of each nuclear experiment.
Nonetheless, in all cases, models conditioned on \EFT~information are favored when we consider all nuclear experiments and astrophysical observations simultaneously (i.e., Bayes factors remain larger than $1$ in Fig.~\ref{fig:evidence}).
We find that the Bayes factors are largest for \MBPT~and \DrischlerPRL~and smallest for \QMC. 
Again, this is likely due to a combination of high-order interactions only present in some calculations, choices of the regulator scheme, and the widths of prior uncertainty bands.

Given the mild tension between the PREX-II value for $R_\mathrm{skin}^{^{208}{\rm Pb}}$ and that inferred from the astrophysical inference with $\chi$EFT information, we investigate what kind of EOS behavior is required to satisfy both the PREX-II and astrophysical constraints.
In Fig.~\ref{fig:pressure-density with L cuts}, we show the pressure and the speed of sound $c_s$ as a function of density for the nonparametric process conditioned only on astrophysical data for all values of $L$, for $30 \mev <L \leq 70\mev$, and for $L> 100\mev$.
Note that this is a stricter requirement than the nominal PREX-II observations suggest at 1-$\sigma$.
We find that the speed of sound generally increases with density.
However, if we assume $L>100$ MeV, we find a local maximum in the median $c_s(n)$ just below $n_0$, although the uncertainties in $c_s$ are large. 
The reason for this feature is that EOSs that are stiff at low densities (large $L$) need to soften beyond $n_0$ to remain consistent with astrophysical data (small tidal deformabilities from GWs). 
Should the PREX-II constraints be confirmed with smaller uncertainty in the future, this might favor the existence of a phase transition between $1$--$2 n_0$. 
However, given current uncertainties, there is no strong preference for such exotic EOS phenomenology based on the data.

\begin{figure}
    \includegraphics[width=0.9\columnwidth, clip=True, trim=0.25cm 2.00cm 0.50cm 4.30cm]{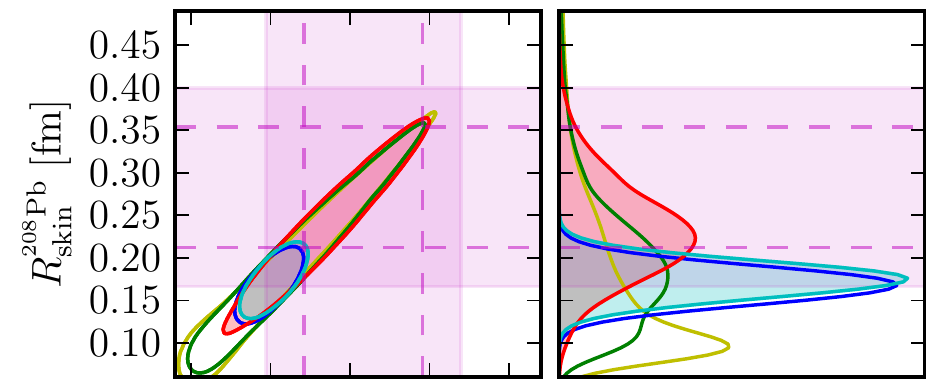} \\
    \includegraphics[width=0.9\columnwidth, clip=True, trim=0.25cm 2.00cm 0.50cm 4.30cm]{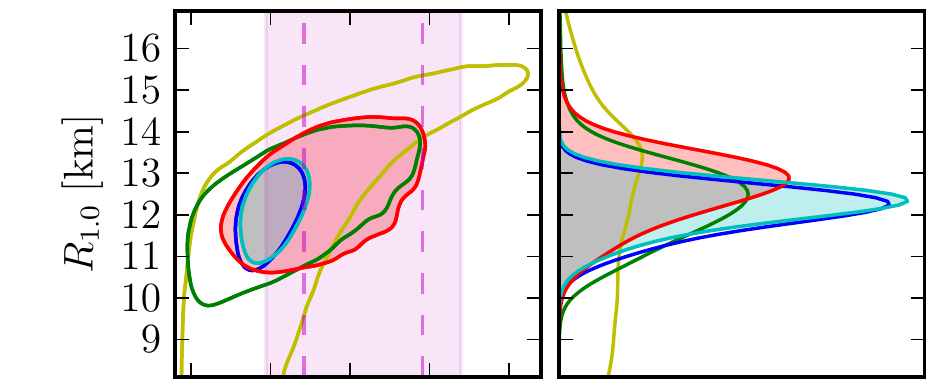} \\
    \includegraphics[width=0.9\columnwidth, clip=True, trim=0.25cm 2.00cm 0.50cm 4.30cm]{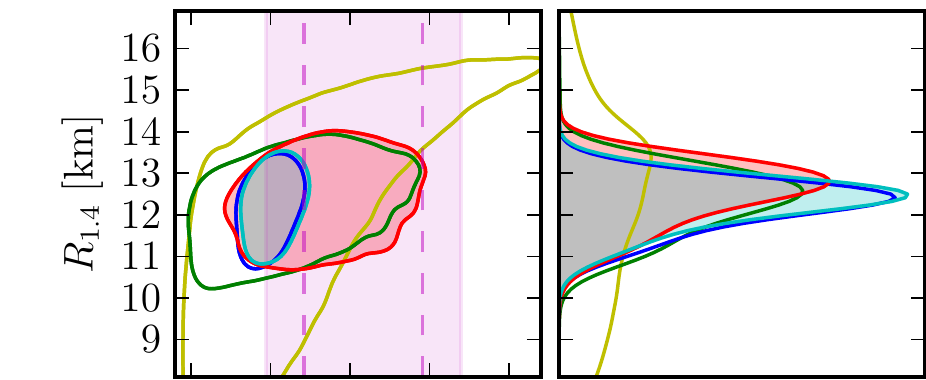} \\
    \includegraphics[width=0.9\columnwidth, clip=True, trim=0.25cm 0.50cm 0.50cm 4.30cm]{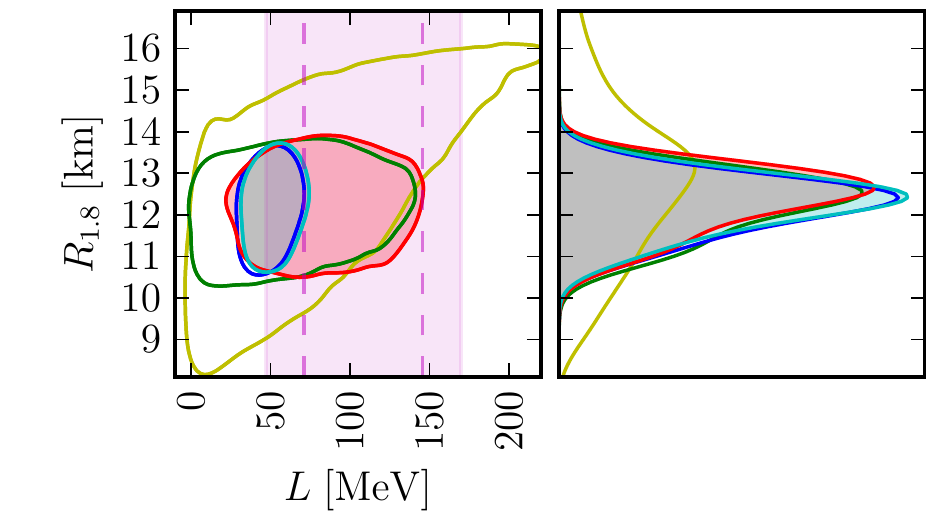} \\
    \caption{
        Correlations of $R_\mathrm{skin}^{^{208}\mathrm{Pb}}$ and the radii of NSs with M=1.0, 1.4, and 1.8 $M_\odot$ with $L$.
        Colors and shading match those in Fig.~\ref{fig:S0 and L}.
    }
    \label{fig:decomposed L-macro correlations}
\end{figure}

Finally, we can ask what would happen to our uncertainty in $S_0$ and $L$ if a series of hypothetical future experiments confirmed the mean of $R_\mathrm{skin}^{^{208}\mathrm{Pb}}$ from PREX-II but with smaller uncertainties.
In Fig.~\ref{fig:pvalue}, we already showed the $p$-values for such a case, which highlight the increased tension with \EFT~calculations.
In Fig.~\ref{fig:S0 and L}, we show the joint posteriors on $S_0$ and $L$ with the current PREX-II uncertainty, half the current uncertainty, and with a perfect $R_\mathrm{skin}^{^{208}\mathrm{Pb}}$ measurement with vanishing uncertainty, where the remaining uncertainty in $L$ is due purely to the uncertainty in the theoretical correlation in Eq.~\eqref{eq:L2Rskin}.
An increased hypothetical precision for $R_\mathrm{skin}^{^{208}\mathrm{Pb}}$ could change our knowledge of $L$ dramatically, possibly rendering it incompatible with the \EFT~predictions when using Eq.~\eqref{eq:L2Rskin}.
However, although the nonparametric Astro+PREX-II posteriors shift compared to the Astro-only posteriors, we never find any significant disagreement.
Indeed, the width of our posterior for $S_0$ is nearly unchanged, even if we assume vanishingly small measurement uncertainty for $R_\mathrm{skin}^{^{208}\mathrm{Pb}}$.
This is another demonstration that current astrophysical data from NSs in the observed mass range cannot strongly constrain nuclear interactions around $n_0$ without further assumptions about the EOS.
The agnostic priors do not closely follow any particular theory (which would generically predict stronger correlations between $S_0$ and $L$).

\subsection{Comparisons between PREX-II, \EFT, and Astrophysical Data for NS Observables}
\label{sec:comparison for NS observables}

Having shown that current strophysical observations of NSs carry only limited information about densities below nuclear saturation, we demonstrate that the inverse is true as well.
Improved measurements of $R_\mathrm{skin}^{^{208}\mathrm{Pb}}$, or even hypothetical direct measurements of $L$, will not significantly improve our knowledge of the macroscopic properties of NSs with masses of $\gtrsim 1.2\,M_\odot$, without additional theory input for the EOS.
Fundamentally, this is because the central densities of astrophysical NSs are above 2$n_0$ (see Ref.~\cite{Legred:2021hdx} for a recent inference of the relation between NS masses and central densities), while the neutron-skin thickness and the symmetry energy parameters describe matter around $n_0$.
Constraints at nuclear saturation density, then, must be extrapolated to higher densities to inform the properties of NSs.
In the nonparametric priors used here, there is enough freedom that such extrapolations only introduce weak correlations between $L$ and, e.g., the radius of NSs.
Strong correlations, like those in Ref.~\cite{Reed:2021nqk}, thus also depend on the model used to describe the EOS above nuclear densities.

\begin{figure}
    \includegraphics[height=0.345\textheight, clip=True, trim=0.0cm 0.5cm 18.4cm 1.0cm]{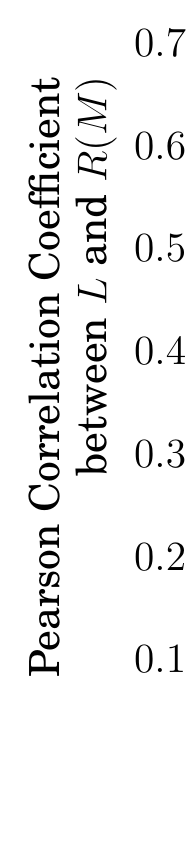}
    \includegraphics[height=0.345\textheight, clip=True, trim=7.95cm 0.5cm 6.30cm 1.0cm]{plot-correlation-coefficients.pdf}
    \caption{
        Pearson correlation coefficients between $L$ and $R(M)$ for marginalized \EFT~results (\emph{blue circles, solid lines}) and \QMC~(\emph{yellow squares, dashed lies}).
        In order from lightest to darkest lines (\emph{top to bottom}), we plot the correlation between $L$ and $R(1.0\,M_\odot)$, $R(1.4\,M_\odot)$, and $R(1.8\,M_\odot)$.
    }
    \label{fig:L_correlations}
\end{figure}

\begin{figure*}
    \begin{minipage}{0.49\textwidth}
    \centering
        \includegraphics[width=0.9\columnwidth, clip=True, trim=0.25cm 2.00cm 0.50cm 4.30cm]{kde-corner-samples_decomposed-L-R_M_1d4_Mondalfit.pdf} \\
        \includegraphics[width=0.9\columnwidth, clip=True, trim=0.25cm 2.00cm 0.50cm 4.30cm]{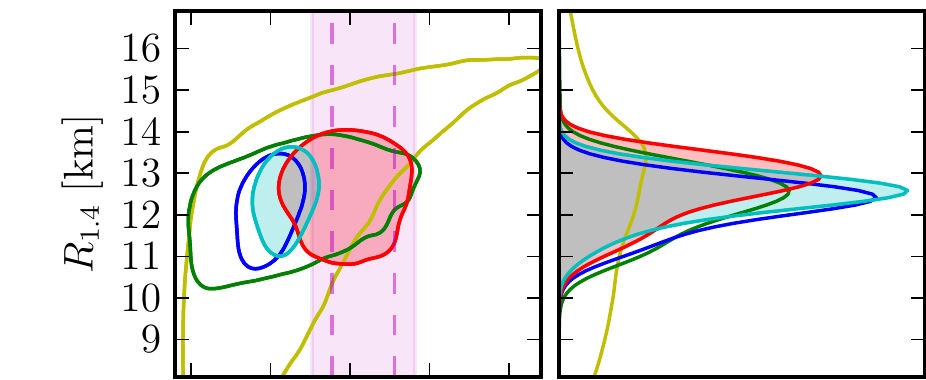} \\
        \includegraphics[width=0.9\columnwidth, clip=True, trim=0.25cm 0.50cm 0.50cm 4.30cm]{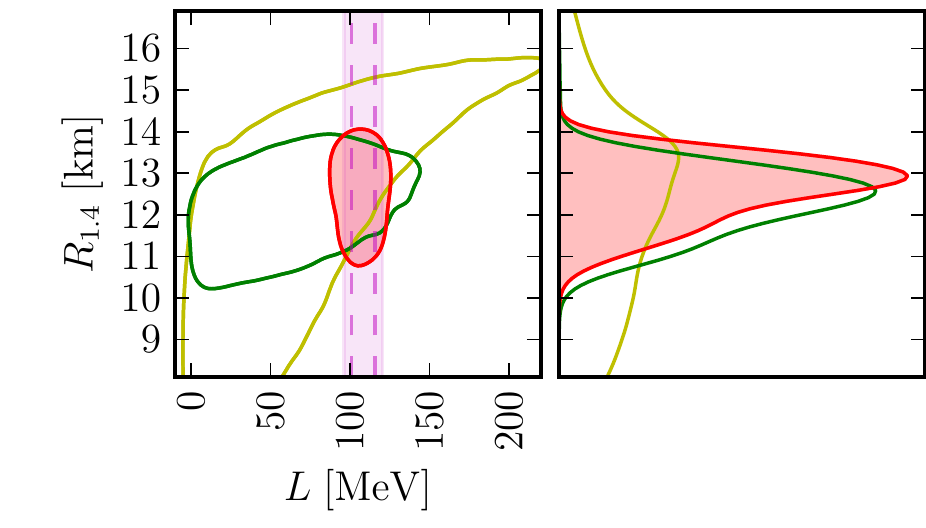} \\
    \end{minipage}
    \begin{minipage}{0.49\textwidth}
        \centering
        \includegraphics[width=0.9\columnwidth, clip=True, trim=0.25cm 2.00cm 0.50cm 4.30cm]{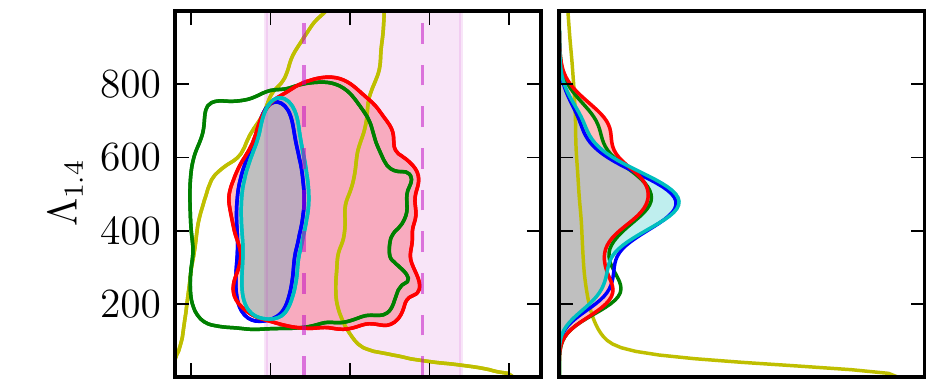} \\
        \includegraphics[width=0.9\columnwidth, clip=True, trim=0.25cm 2.00cm 0.50cm 4.30cm]{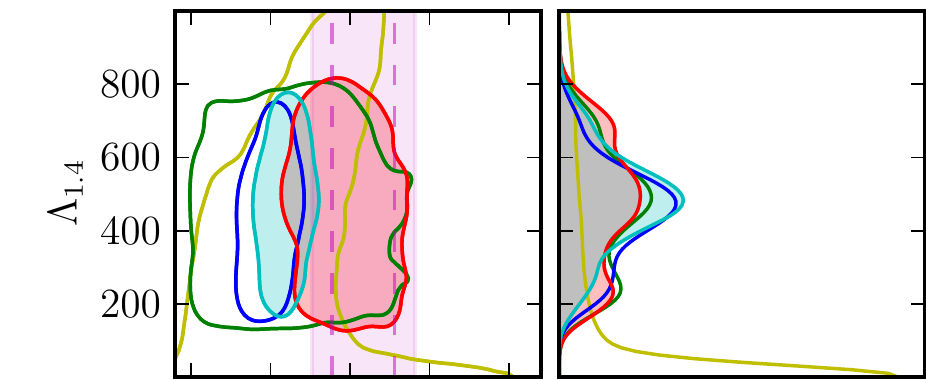} \\
        \includegraphics[width=0.9\columnwidth, clip=True, trim=0.25cm 0.50cm 0.50cm 4.30cm]{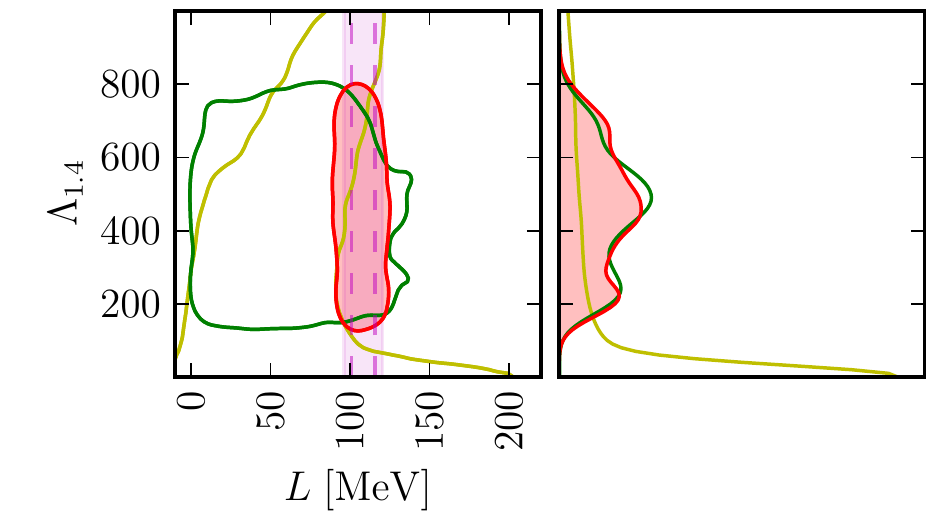} \\
    \end{minipage}
    \caption{
        Correlations between (\emph{left}) $R_{1.4}$, (\emph{right}) $\Lambda_{1.4}$ and $L$ when we model the PREX-II measurement with different uncertainties: the actual measurement uncertainty (\emph{top}), a hypothetical measurement with half the PREX-II uncertainty (\emph{middle}), and a hypothetical measurement with vanishingly small uncertainty for $R_\mathrm{skin}^{^{208}\mathrm{Pb}}$ (\emph{bottom}).
        Colors and shading match those in Fig.~\ref{fig:S0 and L}.
        We see that even a perfect measurement of $R_\mathrm{skin}^{^{208}\mathrm{Pb}}$ does not significantly alter our knowledge of the macroscopic properties of typical astrophysical NSs within the nonparametric inference.
    }
    \label{fig:R(M=1.4), Lambda(M=1.4) and L}
\end{figure*}

We summarize the impact of current $R_\mathrm{skin}^{^{208}\mathrm{Pb}}$ constraints from PREX-II on NSs observables in Fig.~\ref{fig:decomposed L-macro correlations}.
As in Figs.~\ref{fig:violins} and~\ref{fig:S0 and L}, we see that the PREX-II observations do increase the inferred value of $L$ when we do not condition on \EFT.
However, this translates only into a modest shift in the radius of $1.0\,M_\odot$ stars ($R_{1.0}$) and virtually no change for the radii of $1.4$ or $1.8\,M_\odot$ stars ($R_{1.4}$ and $R_{1.8}$, respectively) when we condition on existing astrophysical data.
While we observe correlations between $L$ and $R(M)$ \textit{a priori}, these are intrinsically broad (broader than is often assumed~\cite{Lattimer:2000nx} and \emph{not} one-to-one) and weaken for NSs with higher masses.
These broad correlations are loose enough that astrophysical observations are able to constrain the NS properties while remaining consistent with a wide range of $L$ values.
We find this behavior also in our processes which are conditioned on \EFT~calculations up to $n_0$.

We also consider whether the inclusion of nuclear theory predictions up to higher densities induces stronger correlations between $L$ and $R(M)$ in Fig.~\ref{fig:L_correlations}.
Specifically, we show the Pearson correlation coefficient between $L$ and $R(M)$ under the Astro-only posteriors as a function of the maximum density up to which we trust \EFT.
Generally, we see an increase in the correlation as we trust \EFT~up to higher densities, as expected, although the rate of increase slows at higher densities ($\gtrsim 1.3n_0$) for the \QMC~calculation.
This is likely due to the increase of the theoretical uncertainty band from \QMC~with density, and therefore conditioning on this theoretical prediction imposes a looser constraint. 
Taken to an extreme (high $p_\mathrm{max}$ and small theoretical uncertainties), one sees how trusting a particular theoretical extrapolation to high densities will introduce a strong correlation between $L$ and $R(M)$.
However, we note that the theoretical uncertainties in current \EFT~calculations naturally limit the strength of such correlations, reaching a maximum correlation coefficient of only \result{$\simeq 0.5$} between $L$ and $R_{1.4}$, even when we trust \QMC~up to $> 1.8n_0$.
This may be refined with improved nuclear theory calculations at higher densities.
As expected, the correlation with $L$ is weaker for heavier NSs, see for $R_{1.8}$ in Figs.~\ref{fig:decomposed L-macro correlations} and~\ref{fig:L_correlations}, which is why the recent NICER+XMM observations of J0740+662 ($M=2.08\pm0.07\,M_\odot$)~\cite{Miller:2021qha,Riley:2021pdl} will not constrain the EOS substanially at $n_0$.

Fig.~\ref{fig:R(M=1.4), Lambda(M=1.4) and L} further demonstrates that improved constraints on $L$ will only significantly change our knowledge of $R_{1.4}$ with improved nuclear theory calculations to higher densities (Fig.~\ref{fig:R(M=1.4), Lambda(M=1.4) and L} trusts \EFT~up to $n_0$).
Figure~\ref{fig:R(M=1.4), Lambda(M=1.4) and L} demonstrates this explicitly, where \EFT~input is used only up to $n_0$.
Similar to Fig.~\ref{fig:S0 and L}, we present current constraints on $R_\mathrm{skin}^{^{208}\mathrm{Pb}}$ along with hypothetical measurements with half the uncertainty and with vanishingly small uncertainty for $R_\mathrm{skin}^{^{208}\mathrm{Pb}}$.
Again, while our knowledge of $L$ improves with better measurements of $R_\mathrm{skin}^{^{208}\mathrm{Pb}}$, the inferred posteriors for $R_{1.4}$ and $\Lambda_{1.4}$ are nearly unaffected.\footnote{Reference~\cite{Biswas2021} finds that improved measurements of $R_\mathrm{skin}^{^{208}\mathrm{Pb}}$ can reduce the uncertainty in $R_{1.4}$. We attribute the apparent improvement to correlations introduced by modeling choices made in Ref.~\cite{Biswas2021} (e.g., the extent of the ``low-density'' nuclear parameterization and the polytropic extension to higher densities) that are not introduced within our nonparametric analysis. As elsewhere (e.g.,~\cite{Reed:2021nqk}), reduced uncertainty in NS observables from improved measurements at densities at or below nuclear saturation are contingent upon specific model assumptions that may not be correct.}
In fact, $L$ seems to be particularly uncorrelated with $\Lambda_{1.4}$ within nonparametric extensions, implying that even a perfect measurement of $R_\mathrm{skin}^{^{208}\mathrm{Pb}}$ additionally requires reliable nuclear theory calculations to higher densities to impact our expectations for future GW observations.

\section{Further Discussion}
\label{sec:discussion}

Finally, we discuss possible future areas of improvement and their expected impact, from the assumptions made about the crust EOS, the different neutron matter calculations, translations from pure neutron matter to matter in $\beta$-equilibrium, and the likelihood modeling.
We also briefly discuss additional experimental probes of $R_\mathrm{skin}^{^{208}\mathrm{Pb}}$.

Although we follow the uncertainty of individual \EFT~calculations down to very low densities $n\leq 0.3n_0$, we match all EOS draws to a single BPS crust model~\cite{BaymPethick1971} below that.
Previous work suggested that the uncertainty in the crust at densities below $\simeq 10^{14}\mathrm{g}/\mathrm{cm}^3 = 0.36 n_0$ can lead to a $\leq 0.3\,\mathrm{km}$ change in the radii of typical NSs~\cite{Gamba2019}.
This effect is smaller than our current uncertainty in, e.g., $R_{1.4}$ at the 90\% level (\mrgagnAstroPostROnePointFour~km), but it may not be negligible.
However, our results are qualitatively and quantitatively similar to the results of Ref.~\cite{EssickTewsLandryReddyHolz2020}, which used \EFT~uncertainties down to similarly low densities but connected to a different crust model (SLy~\cite{DouchinHaensel2001}), as well as Refs.~\cite{EssickLandryHolz2020, LandryEssickChatziioannou2020}, which directly marginalized over 3 different crust EOSs (from SLy, ENG~\cite{EngvikOsnes1996} and HQC18~\cite{BaymFurusawa2019}).
Therefore, any uncertainty within the crust model appears to have a minimal impact on our results.

In our work, we explore 4 different \EFT~calculations. These explore \EFT~interactions at different orders, employ different local and nonlocal regularization schemes, and use different many-body methods for the calculation of neutron matter. 
The PNM results are then extended to matter in $\beta$-equilibrium, containing a small fraction of protons and electrons around saturation densities. We emphasize here that for our inference of nuclear matter properties, we focus on densities around $n_0$. This enables the use of expansions around the empirical saturation point. 
These expansions need to be truncated, but this approximation has a negligible effect for the density expansion, again due to the focus on properties at or around $n_0$. In the asymmetry expansion, the truncated higher-order terms beyond $\mathcal{O}(x^2)$ are estimated to be sub-MeV corrections around $n_0$, and can be safely neglected given current EOS uncertainties~\cite{Drischler:2013iza, Somasundaram:2020chb}. 
Nonetheless, this could be improved by future calculations of asymmetric matter around saturation density.

We also note that several approaches to neutron matter calculations and their associated uncertainties exist (see, e.g., discussion in Ref.~\cite{Drischler:2021kxf}).
Our goal in this work was to span a range of different \EFT~calculations instead of attempting to quantify the errors or term-by-term convergence within each individual calculation; thus our choice to marginalize over separate \EFT~estimates.
As such, we took the ``best'' constraint from each calculation instead of, e.g., considering multiple orders within the same calculation (as, e.g., in Ref.~\cite{Drischler:2020fvz}).
While our marginalization renders our conclusions robust and tends to emphasize general trends, future work searching for astrophysical evidence for, e.g. the breakdown scale within \EFT~calculations, will benefit from explicitly checking term-by-term convergence within individual calculations against astrophysical data and further exploring the effects of regulator artefacts.

While our results suggest that higher-order chiral interactions might be important (compare N$^2$LO \QMC~calculations with all other calculations that employ some N$^3$LO contributions) and that locally regularized interactions are less favored (again, compare \QMC~to other calculations) we stress that all \EFT~calculations are consistent with each other and that our conclusions about consistency with nuclear experiment and astrophysical observations apply equally to all four \EFT~calculations. This highlights the robustness of our findings. 

Additionally, one may be concerned with the single-event likelihood models constructed within our hierarchical inference.
We use optimized Gaussian KDEs (see Sec.~\ref{sec:nonparametric gaussian processes}), which have previously been shown to robustly model the associated likelihoods (see, e.g., discussion within Ref.~\cite{EssickLandryHolz2020}).
Indeed, while KDEs are known to be biased approximations to probability densities, these effects are small given the current sample sizes available within public posterior samples for each astrophysical observation we consider.
As Ref.~\cite{EssickLandryHolz2020} discussed, we primarily expect these to impact our estimate of the evidence that a particular object was a BH rather than a NS (due to the sharp boundary at $\Lambda=0$ within GW likelihoods).
We do not consider such an inference here, and therefore expect our KDE models to suffice for the task at hand.
Similar to Refs.~\cite{LandryEssick2019, EssickLandryHolz2020}, we also confirm that we retain large effective numbers of samples throughout all stages of our Monte Carlo inference scheme (typically, $\gtrsim \mathcal{O}(10^4)$ effective EOS samples for our nonparametric and \EFT-marginalized results).
Nevertheless, it is worth noting that other approaches to modeling single-event likelihoods exist in the literature (e.g., Ref.~\cite{Hernandez2019}) which may be of increasing importance with larger numbers of astrophysical observations.

Similarly, marginal likelihoods from astrophysical observations implicitly depend on the mass distributions assumed.
Although the impact of our current assumptions is expected to be small for the existing set of events, larger sample sizes may require simultaneous inference of the NS mass distribution and the EOS, e.g.~\cite{Wysocki:2020myz,Golomb:2021tll}.

Finally, in addition to the approach using weak probes employed by PREX, and the strong correlation with the dipole polarizability from $(p,p')$ scattering, there are other experiments sensitive to $R_\mathrm{skin}^{^{208}\mathrm{Pb}}$ that rely on strong probes, see, e.g., the reviews~\cite{Thiel2019} and~\cite{Tsang:2012se}. While here we have focused on the recent PREX result, and also explored $\alpha_D^{^{208}\mathrm{Pb}}$ due to its well studied strong correlation with $R_\mathrm{skin}^{^{208}\mathrm{Pb}}$, we note that many of the measurements of $R_\mathrm{skin}^{^{208}\mathrm{Pb}}$ that employ strong probes tend to agree more closely with our \EFT~priors, similar to the $\alpha_D^{^{208}\mathrm{Pb}}$ results we consider.
For example, Ref.~\cite{Tarbert2014} estimates $R_\mathrm{skin}^{^{208}\mathrm{Pb}} = 0.15\pm0.03\,(\mathrm{stat.})^{+0.01}_{-0.03}\,(\mathrm{sys.})\,\mathrm{fm}$ based on coherent pion production, and Ref.~\cite{Jastrzebski2004} estimates $0.15\pm0.02\,(\mathrm{stat.})\,\mathrm{fm}$ based on analyses of antiprotonic atoms.
While we do not explicitly consider these in our analysis because of the difficulty in estimating the associated model systematics, future analyses may include them if the model dependence implicit within the experimental results is better understood.

\section{Summary}
\label{sec:summary}

In summary, we used nonparametric EOS inference to constrain the symmetry energy, its density dependence, and $R_\mathrm{skin}^{^{208}\mathrm{Pb}}$ directly from astrophysical data, leading to $S_0 = \mrgagnAstroPostS\,\mev$, $L=\mrgagnAstroPostL\,\mev$, and $R_\mathrm{skin}^{^{208}\mathrm{Pb}}=\mrgagnAstroPostRskin\,\mathrm{fm}$. 
Folding in \EFT~constraints reduces these ranges to $S_0 = \mrgeftAstroPostS\,\mev$, $L=\mrgeftAstroPostL\,\mev$, and $R_\mathrm{skin}^{^{208}\mathrm{Pb}}=\mrgeftAstroPostRskin\,\mathrm{fm}$. 
While these results prefer values below the ones that PREX-II recently reported~\cite{PREXII, Reed:2021nqk}, the PREX-II uncertainties are still broad and any tension is very mild.
Furthermore, our findings are in good agreement with other nuclear physics information.
Our analysis suggests that a future measurement of $R_\mathrm{skin}^{^{208}\mathrm{Pb}}$ with an uncertainty of $\pm 0.04\,\mathrm{fm}$ (a factor of $\simeq 2$ smaller than the current uncertainty) could challenge current \EFT~calculations, although the tension with astrophysical data would still be relatively mild ($p$-value of \mrgagnAstroPostPREXhalfpvalue).
However, we also note that the formation of light clusters at the surface of heavy nuclei could affect the extracted L value~\cite{Tanaka:2021oll}.

Finally, our results demonstrate that the correlation between $R_{1.4}$ and $L$ (or $R_\mathrm{skin}^{^{208}\mathrm{Pb}}$) is looser than suggested by analyses based on a specific class of EOS models.
In fact, even a hypothetically perfect measurements of $R_\mathrm{skin}^{^{208}\mathrm{Pb}}$ will not strongly impact our knowledge of the radius and tidal deformability of $1.4\,M_\odot$ NSs when using nonparametric EOS representations.
The inverse is also true for such EOSs: observations of NSs at astrophysically relevant masses will carry only limited information about nuclear interactions at or below nuclear saturation density. 
Extrapolating neutron-skin thickness measurements to NS scales thus requires a careful treatment of systematic EOS model uncertainties to distinguish implicit modeling assumptions from the data's impact.
In particular, we find that the PREX-II data does not require NSs to have large radii.
However, if the high $L$ values of PREX-II persist, this may suggest a peak in the sound speed around saturation density in order to accommodate both the moderate radii inferred from astrophysical data and the large $L$ observed in terrestrial experiments.
Although tantalizing, it remains to be seen whether astrophysical observations of low-mass NSs or future nuclear experiments will bear this out.

Finally, we note that a confirmation of high values for $S_0$ and $L$ implied by the central PREX-II results would challenge all available microscopic models for nuclear interactions~(see, e.g., Refs.~\cite{Lattimer:2012xj, Tews:2016jhi, Huth:2020ozf, Drischler:2021kxf}). 
This affects both phenomenological two- and three-nucleon potentials as well as interactions derived from \EFT, and would require a significant increase of the repulsion between neutrons at densities of the order of $n_0$. This would have direct implications for studies of the structure of medium-mass to heavy nuclei.


\acknowledgements

R.E. was supported by the Perimeter Institute for Theoretical Physics and the Kavli Institute for Cosmological Physics.
R.E. also thanks the Canadian Institute for Advanced Research (CIFAR) for support.
Research at Perimeter Institute is supported in part by the Government of Canada through the Department of Innovation, Science and Economic Development Canada and by the Province of Ontario through the Ministry of Colleges and Universities.
The Kavli Institute for Cosmological Physics at the University of Chicago is supported by an endowment from the Kavli Foundation and its founder Fred Kavli.
P.L. is supported by National Science Foundation award PHY-1836734 and by a gift from the Dan Black Family Foundation to the Nicholas \& Lee Begovich Center for Gravitational-Wave Physics \& Astronomy.
The work of A.S.~was supported in part by the Deutsche Forschungsgemeinschaft (DFG, German Research Foundation) -- {Project-ID} 279384907 -- SFB 1245.
The work of I.T. was supported by the U.S. Department of Energy, Office of Science, Office of Nuclear Physics, under contract No.~DE-AC52-06NA25396, by the Laboratory Directed Research and Development program of Los Alamos National Laboratory under project numbers 20190617PRD1 and 20190021DR, and by the U.S. Department of Energy, Office of Science, Office of Advanced Scientific Computing Research, Scientific Discovery through Advanced Computing (SciDAC) NUCLEI program.
This work benefited from discussions within IReNA, which is supported in part by the National Science Foundation under Grant No. OISE-1927130.
The authors also gratefully acknowledge the computational resources provided by the LIGO Laboratory and supported by NSF grants PHY-0757058 and PHY-0823459.
Computational resources have also been provided by the Los Alamos National Laboratory Institutional Computing Program, which is supported by the U.S. Department of Energy National Nuclear Security Administration under Contract No.~89233218CNA000001, and by the National Energy Research Scientific Computing Center (NERSC), which is supported by the U.S. Department of Energy, Office of Science, under contract No.~DE-AC02-05CH11231.


\bibliographystyle{apsrev4-1}
\bibliography{biblio}

\end{document}

%% file: EsymAstro_macros-final.tex
\newcommand{\EFT}{\ensuremath{\chi\mathrm{EFT}}}

\newcommand{\QMC}{\ensuremath{\mathrm{QMC}^{(2016)}_\mathrm{N^2LO}}}
\newcommand{\MBPT}{\ensuremath{\mathrm{MBPT}^{(2013)}_\mathrm{N^3LO}}}
\newcommand{\DrischlerPRL}{\ensuremath{\mathrm{MBPT}^{(2019)}_\mathrm{N^3LO}}}
\newcommand{\DrischlerGP}{\ensuremath{\mathrm{MBPT}^{(2020\mathrm{\,GP})}_\mathrm{N^3LO}}}
\newcommand{\Hebeler}{\ensuremath{\mathrm{MBPT}^{(2010)}_\mathrm{mixed}}}

\newcommand{\result}[1]{#1}

\newcommand{\externalresult}[1]{#1}

\newcommand{\prexRskin}{\externalresult{\ensuremath{0.283 \pm 0.071}}}
\newcommand{\prexL}{\externalresult{\ensuremath{106 \pm 37}}}

\newcommand{\mrgagnPriorEpnm }{\result{\ensuremath{17.5^{+14.6}_{-7.7}}}} 
\newcommand{\mrgagnPriorS    }{\result{\ensuremath{33.3^{+14.7}_{-8.2}}}} 
\newcommand{\mrgagnPriorL    }{\result{\ensuremath{38^{+109}_{-41}}}} 
\newcommand{\mrgagnPriorKsym }{\result{\ensuremath{-255^{+853}_{-566}}}} 
\newcommand{\mrgagnPriorRskin}{\result{\ensuremath{0.14^{+0.19}_{-0.09}}}} 
\newcommand{\mrgagnPriorAlphaD}{\result{\ensuremath{18.9^{+4.1}_{-4.7}}}}

\newcommand{\mrgagnAstroPostEpnm }{\result{\ensuremath{19.3^{+11.7}_{-8.5}}}} 
\newcommand{\mrgagnAstroPostS    }{\result{\ensuremath{35.1^{+11.6}_{-8.9}}}} 
\newcommand{\mrgagnAstroPostL    }{\result{\ensuremath{58^{+61}_{-56}}}} 
\newcommand{\mrgagnAstroPostKsym }{\result{\ensuremath{-240^{+559}_{-503}}}} 
\newcommand{\mrgagnAstroPostRskin}{\result{\ensuremath{0.19^{+0.12}_{-0.11}}}} 
\newcommand{\mrgagnAstroPostAlphaD}{\result{\ensuremath{19.0^{+3.8}_{-3.9}}}} 

\newcommand{\mrgagnAstroPostPREXpvalue     }{\result{\ensuremath{25.3\%}}} 
\newcommand{\mrgagnAstroPostPREXhalfpvalue }{\result{\ensuremath{11.5\%}}} 

\newcommand{\mrgagnAstroPostROnePointFour}{\result{\ensuremath{12.39^{+1.02}_{-1.46}}}} 

\newcommand{\mrgagnAstroPREXPostEpnm }{\result{\ensuremath{21.5^{+10.8}_{-8.3}}}} 
\newcommand{\mrgagnAstroPREXPostS    }{\result{\ensuremath{37.3^{+11.8}_{-7.5}}}} 
\newcommand{\mrgagnAstroPREXPostL    }{\result{\ensuremath{80^{+51}_{-46}}}} 
\newcommand{\mrgagnAstroPREXPostKsym }{\result{\ensuremath{-223^{+608}_{-565}}}} 
\newcommand{\mrgagnAstroPREXPostRskin}{\result{\ensuremath{0.23^{+0.10}_{-0.10}}}} 
\newcommand{\mrgagnAstroPREXPostAlphaD}{\result{\ensuremath{19.6^{+3.9}_{-4.4}}}} 

\newcommand{\mrgagnAstroAlphaDPostEpnm }{\result{\ensuremath{18.4^{+7.4}_{-7.8}}}} 
\newcommand{\mrgagnAstroAlphaDPostS    }{\result{\ensuremath{34.2^{+7.4}_{-7.9}}}} 
\newcommand{\mrgagnAstroAlphaDPostL    }{\result{\ensuremath{61^{+49}_{-57}}}} 
\newcommand{\mrgagnAstroAlphaDPostKsym }{\result{\ensuremath{-172^{+483}_{-388}}}} 
\newcommand{\mrgagnAstroAlphaDPostRskin}{\result{\ensuremath{0.19^{+0.10}_{-0.12}}}} 
\newcommand{\mrgagnAstroAlphaDPostAlphaD}{\result{\ensuremath{19.8^{+2.0}_{-2.0}}}} 

\newcommand{\mrgeftPriorEpnm }{\result{\ensuremath{16.7^{+1.5}_{-1.3}}}} 
\newcommand{\mrgeftPriorS    }{\result{\ensuremath{32.5^{+1.9}_{-1.8}}}} 
\newcommand{\mrgeftPriorL    }{\result{\ensuremath{47^{+15}_{-15}}}} 
\newcommand{\mrgeftPriorKsym }{\result{\ensuremath{-119^{+129}_{-133}}}} 
\newcommand{\mrgeftPriorRskin}{\result{\ensuremath{0.16^{+0.04}_{-0.04}}}} 
\newcommand{\mrgeftPriorAlphaD}{\result{\ensuremath{19.6^{+1.7}_{-2.0}}}} 

\newcommand{\mrgeftAstroPostEpnm }{\result{\ensuremath{16.9^{+1.5}_{-1.4}}}} 
\newcommand{\mrgeftAstroPostS    }{\result{\ensuremath{32.7^{+1.9}_{-1.8}}}} 
\newcommand{\mrgeftAstroPostL    }{\result{\ensuremath{49^{+14}_{-15}}}} 
\newcommand{\mrgeftAstroPostKsym }{\result{\ensuremath{-107^{+124}_{-128}}}} 
\newcommand{\mrgeftAstroPostRskin}{\result{\ensuremath{0.17^{+0.04}_{-0.04}}}} 
\newcommand{\mrgeftAstroPostAlphaD}{\result{\ensuremath{19.6^{+1.9}_{-1.7}}}} 

\newcommand{\mrgeftAstroPostPREXpvalue     }{\result{\ensuremath{12.3\%}}} 
\newcommand{\mrgeftAstroPostPREXhalfpvalue }{\result{\ensuremath{ 0.6\%}}} 

\newcommand{\mrgeftAstroPREXPostEpnm }{\result{\ensuremath{17.1^{+1.5}_{-1.5}}}} 
\newcommand{\mrgeftAstroPREXPostS    }{\result{\ensuremath{33.0^{+2.0}_{-1.8}}}} 
\newcommand{\mrgeftAstroPREXPostL    }{\result{\ensuremath{53^{+14}_{-15}}}} 
\newcommand{\mrgeftAstroPREXPostKsym }{\result{\ensuremath{-91^{+118}_{-130}}}} 
\newcommand{\mrgeftAstroPREXPostRskin}{\result{\ensuremath{0.17^{+0.04}_{-0.04}}}} 
\newcommand{\mrgeftAstroPREXPostAlphaD}{\result{\ensuremath{19.8^{+1.7}_{-1.9}}}} 

\newcommand{\mrgeftAstroAlphaDPostEpnm }{\result{\ensuremath{16.9^{+1.5}_{-1.4}}}} 
\newcommand{\mrgeftAstroAlphaDPostS    }{\result{\ensuremath{32.7^{+1.9}_{-1.8}}}} 
\newcommand{\mrgeftAstroAlphaDPostL    }{\result{\ensuremath{51^{+13}_{-14}}}} 
\newcommand{\mrgeftAstroAlphaDPostKsym }{\result{\ensuremath{-98^{+117}_{-124}}}} 
\newcommand{\mrgeftAstroAlphaDPostRskin}{\result{\ensuremath{0.17^{+0.04}_{-0.03}}}} 
\newcommand{\mrgeftAstroAlphaDPostAlphaD}{\result{\ensuremath{19.8^{+1.5}_{-1.9}}}} 

\newcommand{\qmcPriorEpnm }{\result{\ensuremath{16.4^{+1.0}_{-0.9}}}} 
\newcommand{\qmcPriorS    }{\result{\ensuremath{32.2^{+1.5}_{-1.5}}}} 
\newcommand{\qmcPriorL    }{\result{\ensuremath{39^{+11}_{-100}}}} 
\newcommand{\qmcPriorKsym }{\result{\ensuremath{-179^{+111}_{-112}}}} 
\newcommand{\qmcPriorRskin}{\result{\ensuremath{0.15^{+0.03}_{-0.03}}}} 
\newcommand{\qmcPriorAlphaD}{\result{\ensuremath{19.1^{+1.7}_{-1.7}}}} 

\newcommand{\qmcAstroPostEpnm }{\result{\ensuremath{16.5^{+1.1}_{-0.9}}}} 
\newcommand{\qmcAstroPostS    }{\result{\ensuremath{32.4^{+1.5}_{-1.5}}}} 
\newcommand{\qmcAstroPostL    }{\result{\ensuremath{41^{+11}_{-11}}}} 
\newcommand{\qmcAstroPostKsym }{\result{\ensuremath{-165^{+114}_{-112}}}} 
\newcommand{\qmcAstroPostRskin}{\result{\ensuremath{0.15^{+0.03}_{-0.03}}}} 
\newcommand{\qmcAstroPostAlphaD}{\result{\ensuremath{19.2^{+1.6}_{-1.9}}}} 

\newcommand{\qmcAstroPREXPostEpnm }{\result{\ensuremath{16.7^{+1.1}_{-1.0}}}} 
\newcommand{\qmcAstroPREXPostS    }{\result{\ensuremath{32.5^{+1.7}_{-1.4}}}} 
\newcommand{\qmcAstroPREXPostL    }{\result{\ensuremath{44^{+12}_{-12}}}} 
\newcommand{\qmcAstroPREXPostKsym }{\result{\ensuremath{-151^{+124}_{-108}}}} 
\newcommand{\qmcAstroPREXPostRskin}{\result{\ensuremath{0.16^{+0.03}_{-0.03}}}} 
\newcommand{\qmcAstroPREXPostAlphaD}{\result{\ensuremath{19.3^{+1.6}_{-1.9}}}} 

\newcommand{\qmcAstroAlphaDPostEpnm }{\result{\ensuremath{16.5^{+1.1}_{-0.9}}}} 
\newcommand{\qmcAstroAlphaDPostS    }{\result{\ensuremath{32.2^{+1.5}_{-1.5}}}} 
\newcommand{\qmcAstroAlphaDPostL    }{\result{\ensuremath{43^{+11}_{-10}}}} 
\newcommand{\qmcAstroAlphaDPostKsym }{\result{\ensuremath{-153^{+111}_{-107}}}} 
\newcommand{\qmcAstroAlphaDPostRskin}{\result{\ensuremath{0.16^{+0.03}_{-0.03}}}} 
\newcommand{\qmcAstroAlphaDPostAlphaD}{\result{\ensuremath{19.4^{+1.5}_{-1.8}}}} 

\newcommand{\mbptPriorEpnm }{\result{\ensuremath{16.9^{+1.9}_{-1.9}}}} 
\newcommand{\mbptPriorS    }{\result{\ensuremath{32.8^{+2.2}_{-2.2}}}} 
\newcommand{\mbptPriorL    }{\result{\ensuremath{52^{+13}_{-13}}}} 
\newcommand{\mbptPriorKsym }{\result{\ensuremath{-86^{+94}_{-103}}}} 
\newcommand{\mbptPriorRskin}{\result{\ensuremath{0.17^{+0.04}_{-0.03}}}} 
\newcommand{\mbptPriorAlphaD}{\result{\ensuremath{19.9^{+1.6}_{-1.8}}}} 

\newcommand{\mbptAstroPostEpnm }{\result{\ensuremath{17.1^{+1.8}_{-1.9}}}} 
\newcommand{\mbptAstroPostS    }{\result{\ensuremath{32.9^{+2.2}_{-2.1}}}} 
\newcommand{\mbptAstroPostL    }{\result{\ensuremath{53^{+13}_{-12}}}} 
\newcommand{\mbptAstroPostKsym }{\result{\ensuremath{-86^{+96}_{-101}}}} 
\newcommand{\mbptAstroPostRskin}{\result{\ensuremath{0.18^{+0.03}_{-0.04}}}} 
\newcommand{\mbptAstroPostAlphaD}{\result{\ensuremath{19.9^{+1.6}_{-1.8}}}} 

\newcommand{\mbptAstroPREXPostEpnm }{\result{\ensuremath{17.4^{+1.9}_{-1.9}}}} 
\newcommand{\mbptAstroPREXPostS    }{\result{\ensuremath{33.2^{+2.2}_{-2.2}}}} 
\newcommand{\mbptAstroPREXPostL    }{\result{\ensuremath{55^{+13}_{-12}}}} 
\newcommand{\mbptAstroPREXPostKsym }{\result{\ensuremath{-80^{+99}_{-93}}}} 
\newcommand{\mbptAstroPREXPostRskin}{\result{\ensuremath{0.18^{+0.03}_{-0.03}}}} 
\newcommand{\mbptAstroPREXPostAlphaD}{\result{\ensuremath{19.9^{+1.6}_{-1.8}}}} 

\newcommand{\mbptAstroAlphaDPostEpnm }{\result{\ensuremath{17.1^{+1.8}_{-1.9}}}} 
\newcommand{\mbptAstroAlphaDPostS    }{\result{\ensuremath{32.9^{+2.1}_{-2.0}}}} 
\newcommand{\mbptAstroAlphaDPostL    }{\result{\ensuremath{54^{+13}_{-12}}}} 
\newcommand{\mbptAstroAlphaDPostKsym }{\result{\ensuremath{-84^{+102}_{-92}}}} 
\newcommand{\mbptAstroAlphaDPostRskin}{\result{\ensuremath{0.18^{+0.03}_{-0.04}}}} 
\newcommand{\mbptAstroAlphaDPostAlphaD}{\result{\ensuremath{19.9^{+1.5}_{-1.8}}}} 

\newcommand{\DrischlerPRLPriorEpnm }{\result{\ensuremath{17.0^{+1.4}_{-1.4}}}} 
\newcommand{\DrischlerPRLPriorS    }{\result{\ensuremath{32.8^{+1.8}_{-1.8}}}} 
\newcommand{\DrischlerPRLPriorL    }{\result{\ensuremath{53^{+12}_{-12}}}} 
\newcommand{\DrischlerPRLPriorKsym }{\result{\ensuremath{-63^{+117}_{-113}}}} 
\newcommand{\DrischlerPRLPriorRskin}{\result{\ensuremath{0.18^{+0.03}_{-0.03}}}} 
\newcommand{\DrischlerPRLPriorAlphaD}{\result{\ensuremath{20.0^{+1.6}_{-1.9}}}} 

\newcommand{\DrischlerPRLAstroPostEpnm }{\result{\ensuremath{17.1^{+1.3}_{-1.2}}}} 
\newcommand{\DrischlerPRLAstroPostS    }{\result{\ensuremath{32.9^{+1.8}_{-1.7}}}} 
\newcommand{\DrischlerPRLAstroPostL    }{\result{\ensuremath{54^{+11}_{-11}}}} 
\newcommand{\DrischlerPRLAstroPostKsym }{\result{\ensuremath{-63^{+114}_{-117}}}} 
\newcommand{\DrischlerPRLAstroPostRskin}{\result{\ensuremath{0.18^{+0.03}_{-0.03}}}} 
\newcommand{\DrischlerPRLAstroPostAlphaD}{\result{\ensuremath{20.0^{+1.6}_{-1.9}}}} 

\newcommand{\DrischlerPRLAstroPREXPostEpnm }{\result{\ensuremath{17.2^{+1.3}_{-1.3}}}} 
\newcommand{\DrischlerPRLAstroPREXPostS    }{\result{\ensuremath{33.1^{+1.7}_{-1.8}}}} 
\newcommand{\DrischlerPRLAstroPREXPostL    }{\result{\ensuremath{56^{+11}_{-12}}}} 
\newcommand{\DrischlerPRLAstroPREXPostKsym }{\result{\ensuremath{-53^{+115}_{-116}}}} 
\newcommand{\DrischlerPRLAstroPREXPostRskin}{\result{\ensuremath{0.18^{+0.03}_{-0.03}}}} 
\newcommand{\DrischlerPRLAstroPREXPostAlphaD}{\result{\ensuremath{20.1^{+1.5}_{-2.0}}}} 

\newcommand{\DrischlerPRLAstroAlphaDPostEpnm }{\result{\ensuremath{17.1^{+1.3}_{-1.3}}}} 
\newcommand{\DrischlerPRLAstroAlphaDPostS    }{\result{\ensuremath{32.9^{+1.7}_{-1.6}}}} 
\newcommand{\DrischlerPRLAstroAlphaDPostL    }{\result{\ensuremath{54^{+11}_{-11}}}} 
\newcommand{\DrischlerPRLAstroAlphaDPostKsym }{\result{\ensuremath{-61^{+111}_{-114}}}} 
\newcommand{\DrischlerPRLAstroAlphaDPostRskin}{\result{\ensuremath{0.18^{+0.03}_{-0.03}}}} 
\newcommand{\DrischlerPRLAstroAlphaDPostAlphaD}{\result{\ensuremath{20.0^{+1.5}_{-1.8}}}} 

\newcommand{\DrischlerGPPriorEpnm }{\result{\ensuremath{16.9^{+1.2}_{-1.2}}}} 
\newcommand{\DrischlerGPPriorS    }{\result{\ensuremath{32.8^{+1.7}_{-1.7}}}} 
\newcommand{\DrischlerGPPriorL    }{\result{\ensuremath{53^{+10}_{-10}}}} 
\newcommand{\DrischlerGPPriorKsym }{\result{\ensuremath{-87^{+99}_{-101}}}} 
\newcommand{\DrischlerGPPriorRskin}{\result{\ensuremath{0.17^{+0.03}_{-0.03}}}} 
\newcommand{\DrischlerGPPriorAlphaD}{\result{\ensuremath{20.0^{+1.5}_{-1.9}}}} 

\newcommand{\DrischlerGPAstroPostEpnm }{\result{\ensuremath{17.0^{+1.3}_{-1.1}}}} 
\newcommand{\DrischlerGPAstroPostS    }{\result{\ensuremath{32.8^{+1.7}_{-1.5}}}} 
\newcommand{\DrischlerGPAstroPostL    }{\result{\ensuremath{53^{+9}_{-10}}}} 
\newcommand{\DrischlerGPAstroPostKsym }{\result{\ensuremath{-86^{+95}_{-104}}}} 
\newcommand{\DrischlerGPAstroPostRskin}{\result{\ensuremath{0.18^{+0.03}_{-0.03}}}} 
\newcommand{\DrischlerGPAstroPostAlphaD}{\result{\ensuremath{20.0^{+1.5}_{-1.9}}}} 

\newcommand{\DrischlerGPAstroPREXPostEpnm }{\result{\ensuremath{17.1^{+1.2}_{-1.1}}}} 
\newcommand{\DrischlerGPAstroPREXPostS    }{\result{\ensuremath{32.9^{+1.7}_{-1.6}}}} 
\newcommand{\DrischlerGPAstroPREXPostL    }{\result{\ensuremath{54^{+10}_{-9}}}} 
\newcommand{\DrischlerGPAstroPREXPostKsym }{\result{\ensuremath{-81^{+98}_{-97}}}} 
\newcommand{\DrischlerGPAstroPREXPostRskin}{\result{\ensuremath{0.18^{+0.03}_{-0.03}}}} 
\newcommand{\DrischlerGPAstroPREXPostAlphaD}{\result{\ensuremath{20.0^{+1.5}_{-1.9}}}} 

\newcommand{\DrischlerGPAstroAlphaDPostEpnm }{\result{\ensuremath{17.0^{+1.3}_{-1.1}}}} 
\newcommand{\DrischlerGPAstroAlphaDPostS    }{\result{\ensuremath{32.8^{+1.7}_{-1.4}}}} 
\newcommand{\DrischlerGPAstroAlphaDPostL    }{\result{\ensuremath{53^{+10}_{-9}}}} 
\newcommand{\DrischlerGPAstroAlphaDPostKsym }{\result{\ensuremath{-85^{+93}_{-103}}}} 
\newcommand{\DrischlerGPAstroAlphaDPostRskin}{\result{\ensuremath{0.18^{+0.03}_{-0.03}}}} 
\newcommand{\DrischlerGPAstroAlphaDPostAlphaD}{\result{\ensuremath{20.0^{+1.4}_{-1.9}}}} 

\newcommand{\HebelerPriorEpnm }{\result{\ensuremath{16.6^{+1.2}_{-1.2}}}} 
\newcommand{\HebelerPriorS    }{\result{\ensuremath{32.4^{+1.7}_{-1.6}}}} 
\newcommand{\HebelerPriorL    }{\result{\ensuremath{43^{+11}_{-11}}}} 
\newcommand{\HebelerPriorKsym }{\result{\ensuremath{-149^{+104}_{-100}}}} 
\newcommand{\HebelerPriorRskin}{\result{\ensuremath{0.16^{+0.03}_{-0.03}}}} 
\newcommand{\HebelerPriorAlphaD}{\result{\ensuremath{19.3^{+1.7}_{-1.7}}}} 

\newcommand{\HebelerAstroPostEpnm }{\result{\ensuremath{16.7^{+1.3}_{-1.2}}}} 
\newcommand{\HebelerAstroPostS    }{\result{\ensuremath{32.6^{+1.7}_{-1.7}}}} 
\newcommand{\HebelerAstroPostL    }{\result{\ensuremath{44^{+12}_{-11}}}} 
\newcommand{\HebelerAstroPostKsym }{\result{\ensuremath{-145^{+101}_{-103}}}} 
\newcommand{\HebelerAstroPostRskin}{\result{\ensuremath{0.16^{+0.03}_{-0.03}}}} 
\newcommand{\HebelerAstroPostAlphaD}{\result{\ensuremath{19.3^{+1.7}_{-1.7}}}} 

\newcommand{\HebelerAstroPREXPostEpnm }{\result{\ensuremath{16.9^{+1.3}_{-1.3}}}} 
\newcommand{\HebelerAstroPREXPostS    }{\result{\ensuremath{32.8^{+1.8}_{-1.7}}}} 
\newcommand{\HebelerAstroPREXPostL    }{\result{\ensuremath{47^{+12}_{-12}}}} 
\newcommand{\HebelerAstroPREXPostKsym }{\result{\ensuremath{-138^{+100}_{-102}}}} 
\newcommand{\HebelerAstroPREXPostRskin}{\result{\ensuremath{0.16^{+0.03}_{-0.04}}}} 
\newcommand{\HebelerAstroPREXPostAlphaD}{\result{\ensuremath{19.4^{+1.6}_{-1.8}}}} 

\newcommand{\HebelerAstroAlphaDPostEpnm }{\result{\ensuremath{16.7^{+1.3}_{-1.3}}}} 
\newcommand{\HebelerAstroAlphaDPostS    }{\result{\ensuremath{32.5^{+1.7}_{-1.7}}}} 
\newcommand{\HebelerAstroAlphaDPostL    }{\result{\ensuremath{46^{+12}_{-11}}}} 
\newcommand{\HebelerAstroAlphaDPostKsym }{\result{\ensuremath{-138^{+97}_{-101}}}} 
\newcommand{\HebelerAstroAlphaDPostRskin}{\result{\ensuremath{0.16^{+0.03}_{-0.03}}}} 
\newcommand{\HebelerAstroAlphaDPostAlphaD}{\result{\ensuremath{19.5^{+1.5}_{-1.8}}}} 